\documentclass[aps,prl,twocolumn,superscriptaddress,groupedaddress
 amsmath,amssymb,
footinbib,
 reprint,%
]{revtex4-1}

\usepackage{graphicx}% Include figure files
\usepackage{dcolumn}% Align table columns on decimal point
\usepackage{bm}% bold math
\usepackage[utf8]{inputenc}
\usepackage[T1]{fontenc}
\usepackage{mathptmx}
\usepackage{textgreek}

\usepackage{color}

\usepackage{natbib}

\newcommand{\beginsupplement}{%
        \setcounter{table}{0}
        \renewcommand{\thetable}{S\arabic{table}}%
        \setcounter{figure}{0}
        \renewcommand{\thefigure}{S\arabic{figure}}%
				\renewcommand{\theequation}{S.\arabic{equation}}
     }

\begin{document}

\title[JJ]{Josephson inductance as a probe for highly ballistic semiconductor-superconductor weak links} 
% Force line breaks with \\

\author{Christian Baumgartner}
\author{Lorenz Fuchs}
\author{Linus Fr\'esz} %Frész }
\author{Simon Reinhardt}
\affiliation{Institut f\"ur Experimentelle und Angewandte Physik, University of Regensburg, 93053 Regensburg, Germany}
\author{Sergei Gronin}
\author{Geoffrey C.~Gardner}
\affiliation{Microsoft Quantum Purdue, Purdue University, West Lafayette, Indiana 47907 USA}
\affiliation{Birck Nanotechnology Center, Purdue University, West Lafayette, Indiana 47907 USA}
\author{Michael J.~Manfra}
\affiliation{Department of Physics and Astronomy, Purdue University, West Lafayette, Indiana 47907 USA}
\affiliation{Birck Nanotechnology Center, Purdue University, West Lafayette, Indiana 47907 USA}
\affiliation{School of Materials Engineering, Purdue University, West Lafayette, Indiana 47907 USA}
\affiliation{School of Electrical and Computer Engineering, Purdue University, West Lafayette, Indiana 47907 USA}
\affiliation{Microsoft Quantum Purdue, Purdue University, West Lafayette, Indiana 47907 USA}

\author{Nicola Paradiso}\email{nicola.paradiso@physik.uni-regensburg.de}
\author{Christoph Strunk}
\affiliation{Institut f\"ur Experimentelle und Angewandte Physik, University of Regensburg, 93053 Regensburg, Germany}

\begin{abstract}	
We present simultaneous measurements of Josephson inductance and DC transport characteristics of ballistic Josephson junctions based upon an epitaxial Al-InAs heterostructure. The Josephson inductance at finite current bias directly reveals the current-phase relation. The proximity-induced gap, the critical current and the average value of the transparency $\bar{\tau}$ are extracted without need for phase bias, demonstrating, e.g.,~a near-unity value of $\bar{\tau}=0.94$. Our method allows us to probe the devices deeply in the non-dissipative regime, where ordinary transport measurements are featureless. In perpendicular magnetic field the junctions show a nearly perfect Fraunhofer pattern of the critical current, which is insensitive to the value of $\bar{\tau}$.  In contrast, the signature of supercurrent interference in the inductance turns out to be extremely sensitive to $\bar{\tau}$. 
\end{abstract}

\maketitle 

Epitaxial semiconductor-superconductor hybrids~\cite{Krogstrup_2015,Chang_2015} have 
provided an important platform for new types of devices including basic elements for topological quantum computing~\cite{Shabani2016}.
The epitaxial growth enabled a new generation of proximity-coupled Josephson junctions (JJs) that constitutes an unique playground in modern condensed matter physics research. In such junctions, the relation $I(\varphi)$ between supercurrent $I$ and phase difference $\varphi$ between superconducting leads encodes information on the rich physics of Andreev bound states (ABS) \cite{Furusaki_1991,BeenakkerPRL91,DellaRocca2007}.  Particularly exciting phenomena emerge in the presence of strong spin-orbit interaction as, e.g., for InAs-based junctions~\cite{deVries_Manfra_CM_LK2018, Ke_Goswami_2019,Mayer2019}.  Topologically protected phases have been predicted~\cite{Alicea_2012,Hell2017,Pientka2017,Scharf2019}  and recently demonstrated~\cite{Ren2019,Fornieri2019}. Moreover, simultaneous breaking of both time-reversal and parity symmetry~\cite{Rasmussen2016} leads to an anomalous shift in the current phase relation~\cite{Reynoso_2008,Yokoyama_Eto_Nazarov_2014,Szombati_2016,Mayer2020}, so that the junctions exhibit finite phase difference at zero-current, and vice versa.

Current-voltage [$I(V)$] characteristics of single junctions are simple to measure, but do not provide access to the current-phase relation (CPR). Typically, an asymmetric SQUID \cite{Delagrange2015,Murani2017,ChuanLi_Brinkman2018}, or a local probe of the magnetic field  \cite{Fuechsle2009, Spanton_2017,Hart_Moler_2019}  is needed to implement the phase bias. Alternatively, the phase dependence of the Josephson inductance $L(\varphi)=[(2\pi/\Phi_0)\cdot dI(\varphi)/d\varphi]^{
-1}$ has been measured using a superconducting microwave resonator \cite{Dassonneville2013,Murani2019,Tosi_Pothier_ABS_2019}. However, such resonators are usually not compatible with high magnetic fields. Another option is the interferometer-based method described in Ref.~\cite{RMPGolubov} which, however, does not provide access to the DC transport properties. On the other hand, it should be possible to investigate the non-linear inductance $L(I)$ that is obtained by eliminating the unknown phase from the two equations $I(\varphi)$ and $L^{-1}(\varphi)$.  This route seems so far nearly unexplored in the context of proximity-coupled JJs.
In addition, measurements of individual multichannel junctions are always affected by the sample-specific defect configuration, which tends to mask the underlying generic properties of the specific semiconductor material. Hence a method is desirable, which provides an average over a large ensemble of junctions, in which the effects of individual defect configurations have negligible effect.

\begin{figure}[tb]
\includegraphics[width=\columnwidth]{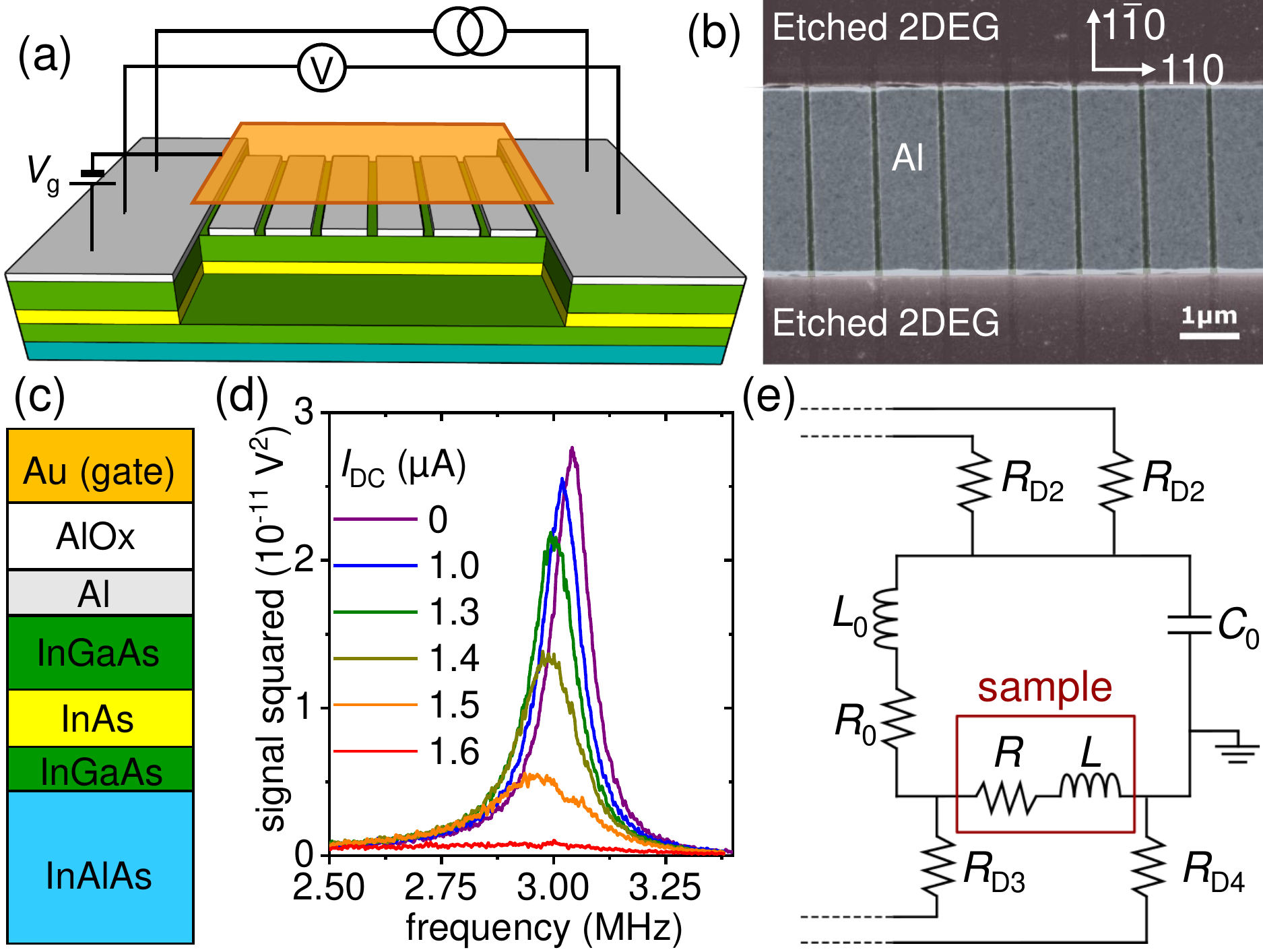}
\caption{(a) Schematic of the Josephson junction array. The actual array is made of 2250 Al islands. (b) Scanning electron micrograph of a portion of the array, taken prior to the deposition of the gate dielectric and of the global top-gate. (c) Sequence of the topmost layers for the heterostructure under study. The Al oxide and the Au layer have been lithographically deposited after the wafer growth.   (d) RLC resonance spectra for different values of the DC current through the array of Josephson junctions, measured at $T=500$~mK.  (e) Circuit scheme of the cold RLC resonator used in this work. 
}
\label{fig:firstfig}
\end{figure}

In this Letter we report on both the Josephson inductance and the DC transport characteristics of a linear array of about 2250 individual junctions. 
We show that the dependence of the Josephson inductance on current bias, magnetic field and temperature is quantitatively understood in terms of the short ballistic junction model. From the data, we deduce an average transparency very close to one. We infer also the induced superconducting gap and the number of channels carrying the supercurrent.  As opposed to the critical current, the quantum interference pattern in the inductance is very sensitive to the transparency. We find perfect consistency between the DC-current and magnetic field dependence of the inductance. Our method provides a simple,  versatile and robust access to the ABS physics in multi-channel unconventional Josephson junctions.

\begin{figure*}[tb]
\includegraphics[width=2\columnwidth]{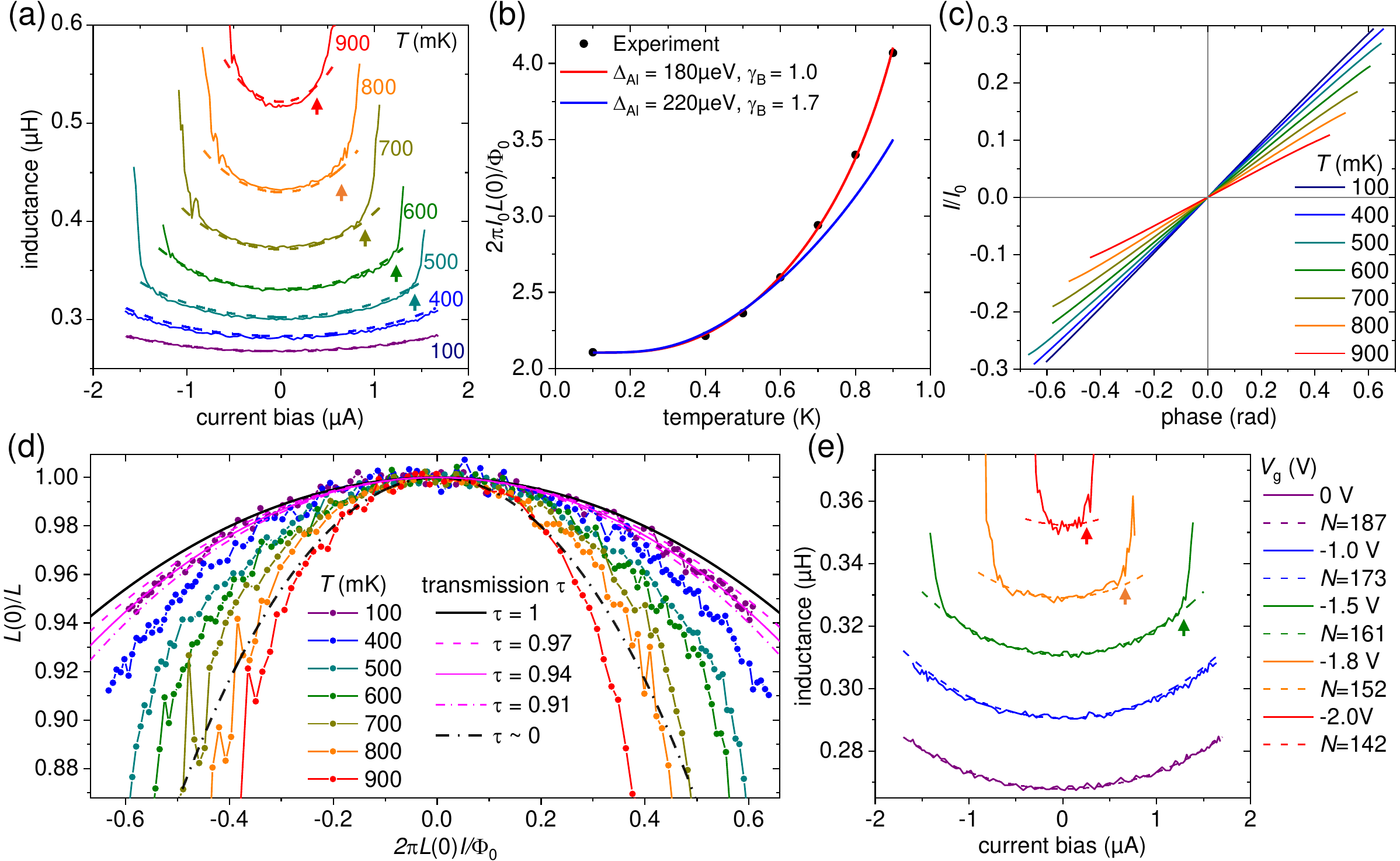}
\caption{ (a) Josephson inductance $L$ versus current bias $I$ measured for different temperatures from 100 to 900~mK (solid lines). Dashed lines show $L$ computed from Eq.~\ref{eq:cprshortball} with parameters $I_0=5.882$~\textmu A, $\bar{\tau}=0.94$, $\Delta_{\text{Al}}=180$~\textmu eV and $\gamma_B=1.0$ (see text). (b) Zero-bias inductance $L(0)$, normalized to $\Phi_0/(2\pi I_0)$, plotted versus temperature (symbols), together with the prediction from  Eq.~\ref{eq:cprshortball} for (red curve) $\Delta_{\text{Al}}=180$~\textmu eV, $\gamma_B=1.0$ and (blue curve) $\Delta_{\text{Al}}=220$~\textmu eV, $\gamma_B=1.7$.    (c) CPR curves obtained by integrating data in panel (a) using Eq.~\ref{eq:inversecpr}. (d) Symbols show a normalized representation of data in panel (a) (see text). Lines show the prediction of the $T=0$ limit of Eq.~\ref{eq:cprshortball} for selected values of the transparency $\bar{\tau}$. (e) $L(I)$ at $T=100$~mK plotted for different gate voltage values $V_{g}$ (solid lines), together with the computed $L(I)$ from Eq.~\ref{eq:cprshortball} (dashed lines). The number  of supercurrent-carrying channels is deduced from the number $N$ in Eq.~\ref{eq:I0} that best fits the data.} 
\label{fig:LIDC}
\end{figure*}

Our samples are fabricated starting from a heterostructure based on a 7~nm-thick Al film epitaxially grown on top of a InGaAs/InAs quantum well [see Fig.~\ref{fig:firstfig}(a,c)], producing a shallow 2D electron gas (2DEG)~\footnote{Further information is provided in the Supplementary Material.}.
The whole array is covered with a 40~nm-thick aluminum oxide layer and with a  5~nm Ti/120~nm Au metal film used as a global top-gate.
The 2DEG underneath the epitaxial Al film is proximitized, with an induced gap $\Delta^{\ast}\approx 140$~\textmu eV, as determined by tunnel spectroscopy~\cite{Note1} using a quantum point contact prepared on a separate chip from the same wafer (similar as in Ref.~\cite{Kjaergaard2016}). 
A JJ array of about 2250 islands is produced by standard lithographic techniques. The island width, length and separation is 3.15, 1.0 and 0.10~\textmu m, respectively. 
The Josephson inductance of such a large number of junctions in series produces a sizable total inductance, of the order of hundreds of nH. The differential inductance $L(I)$ is inferred from the resonance frequency shift~\cite{Meservey1969} [see Fig.~\ref{fig:firstfig}(d)] of a cold RLC circuit, sketched in Fig.~\ref{fig:firstfig}(e), mounted directly on the sample holder~\cite{Note1}. The external inductor (capacitor) has an inductance (capacitance)  $L_0=382$~nH ($C_0=4$~nF). Figure~\ref{fig:firstfig}(d) shows typical resonance spectra for different values of the DC current bias at 500~mK.  By automated fitting, we extract the center frequency and thus the array inductance $L$, which is reported in what follows.

The Josephson inductance is computed starting from the time-derivative of the CPR $I=I_0 f(\varphi)$, where $I_0$ is the characteristic current scale~\footnote{If $\max |f(\varphi)|=1$, then $I_0$ is the critical current of the junction. }, $\varphi$ is the phase difference between the superconducting leads and $f(\varphi)$ a $2\pi$-periodic dimensionless function  (e.g., $f(\varphi)=\sin \varphi$ for a tunnel junction). The ratio of Josephson voltage $V=\hbar\dot{\varphi}/2e$ and time-derivative of the CPR defines the Josephson inductance
\begin{equation}
L(\varphi)\ \equiv\ \frac{V}{\frac{dI}{dt}}\ =\ \frac {\Phi_0}{2\pi I_0f'(\varphi)}.
\label{eq:JI}
\end{equation}
Integration of $L\dot{I}=\Phi_0\dot{\varphi}/2\pi$ provides a reconstruction of the (inverse) CPR $\varphi=\varphi(I)$:
\begin{equation}
\varphi(I)\ =\ \varphi(0)+\frac{2\pi}{\Phi_0}\,\int_0^I L(I')dI',
\label{eq:inversecpr}
\end{equation}
where $L(I)$ is the measured inductance as a function of the DC current bias. We stress that here the phase difference is controlled by the current bias, as opposed to the asymmetric SQUID method where $\varphi$ is controlled by the magnetic flux in the loop.

Solid lines in Fig.~\ref{fig:LIDC}(a) show the Josephson inductance measured as a function of current bias at different temperatures. We notice that an increase of temperature produces an increase of the zero-bias inductance $L(0)$. This is further increased by a finite current bias. In order to quantitatively describe our data, we made use of the CPR for short ballistic junctions at arbitrary temperature, which is given by~\cite{BeenakkerPRL91,DellaRocca2007,RMPGolubov}
\begin{equation}
I(\varphi)=I_0f(\varphi)=I_0\frac{\bar{\tau}\sin \varphi \tanh \left[ \frac{\Delta^{\ast}(T)}{2k_BT}\sqrt{1-\bar{\tau}\sin^2\left(\frac{\varphi}{2}\right)} \right]}{2\sqrt{1-\bar{\tau}\sin^2\left(\frac{\varphi}{2}\right)}},
\label{eq:cprshortball}
\end{equation}
 where $\Delta^{\ast}(T)$ is the induced superconducting gap of the proximitized 2DEG and $\bar{\tau}$ is an average transmission coefficient~\cite{Note1}. Note that $I_0$ corresponds to the critical current only for $\bar{\tau}=1$ and $T=0$. We shall show that all our results are very well described by Eq.~\ref{eq:cprshortball}, even though our 2250 junctions are in the multichannel regime. 
The accessible part of the CPR $I(\varphi)$ corresponding to the data in Fig.~\ref{fig:LIDC}(a) is obtained using Eq.~\ref{eq:inversecpr} and plotted in Fig.~\ref{fig:LIDC}(c). 
In order to better compare the current dependence of the curves in Fig.~\ref{fig:LIDC}(a) with that expected from Eq.~\ref{eq:cprshortball}, we plotted them in a normalized form in Fig.~\ref{fig:LIDC}(d). 
This graph shows $L(0)/L(I)$ plotted as a function of $2\pi L(0)I/\Phi_0$. 
This normalization allows us to express the results in a form that is sensitive only to the shape of the CPR (i.e.,~to $\bar{\tau}$) and not to its prefactor $I_0$. In  Fig.~\ref{fig:LIDC}(d) we observe that an increase of temperature produces an increase of curvature for $L(I)$. The solid and dash-dotted black lines represent the limiting cases for $\bar{\tau}\rightarrow 1$ and $\bar{\tau}\rightarrow 0$ in Eq.~\ref{eq:cprshortball}, respectively. The lowest temperature curve ($T=100$~mK) matches with $\bar{\tau}=0.94$. The other important parameter $I_0=5.882$~\textmu A is then obtained from the $L(0)$  value at the same temperature  using Eq. (1) with $\bar{tau}=0.94$ in the function $f$. This corresponds to a critical current $I_c \equiv  I_0\max_{\varphi}f (\bar{\tau}=0.94, \varphi) = 4.41$~\textmu A, which is about 0.75$I_0$.

The temperature dependence of the Josephson inductance provides the induced gap $\Delta^{\ast}$. By fitting the measured values of $L(0)$ versus $T$, shown in  Fig.~\ref{fig:LIDC}(b), it is possible to extract the last two parameters of our problem, namely the Al gap $\Delta_{\text{Al}}$ and the barrier parameter $\gamma_B$ between Al film and 2DEG. As discussed in Ref.~\cite{KjaergaardPRAPPL17}, these two parameters determine~\cite{Chrestin1997,Aminov1996,Schaepersbook} the temperature dependence of the induced gap $\Delta^{\ast}$ (see discussion in the Supplementary Material~\cite{Note1}). The fit in Fig.~\ref{fig:LIDC}(b) (red line) provides the values $\Delta_{\text{Al}}=180$~\textmu eV and $\gamma_B=1.0$. Alternatively, $\Delta_{\text{Al}}$ can be estimated from $T_c$~\cite{Note1}, leaving $\gamma_B$ as the only fitting parameter. In this case the fit [blue line in Fig.~\ref{fig:LIDC}(b)] underestimates $L(0)$ at higher temperature. In both cases, we obtain $\Delta^{\ast}(0)\approx 130$~\textmu eV, in agreement with the value found in tunneling data mentioned above.

Inserting the four parameters $\bar{\tau}=0.94$, $I_0=5.882$~\textmu A, $\Delta_{\text{Al}}=180$~\textmu eV and $\gamma_B=1.0$ just determined into Eq.~\ref{eq:cprshortball}, we obtain a consistent quantitative description of our whole set of data. We begin with Fig.~\ref{fig:LIDC}(a). Without adjustment, the dashed lines perfectly match the curvature of $L(I)$ up to  the appearance of the upwards kinks, marked with arrows. These kinks correspond to some weak junctions in the array with reduced critical current. At moderate bias their inductance is negligible compared to that of the other two thousand junctions in series. However, when the current approaches their reduced critical current value, their inductance sharply increases until it becomes dominant. At the same time, the resistance quickly increases and damps out the resonance~\cite{Note1}. The kinks become discontinuities at the lowest temperatures [resonance damped within one experimental point in Fig.~\ref{fig:LIDC}(a)], indicating that there are only a few of such weaker junctions. Their reduced critical current sets the highest current at which the inductance can be measured, which is markedly less than $I_0$ found \textit{at equilibrium} with inductance measurements. This limits the accessible fraction of the CPR as shown in Fig.~\ref{fig:LIDC}(c).
We stress that, while dominating the transport at high bias, weak junctions are irrelevant at moderate bias, provided the array is long enough. The larger the number of junctions, the less important are imperfections in few of them. 

Once the relevant CPR parameters have been found, it is possible to further validate our analysis by investigating the dependence of $L$ on other parameters. Figure~\ref{fig:LIDC}(e) shows how the measured finite-bias $L(I)$ (solid lines) depends on the gate voltage $V_g$. At a first glance, the curves resemble those in Fig.~\ref{fig:LIDC}(a), i.e., $L$ increases by increasing $|V_g|$ and $|I|$. There is, however, an important difference: in Fig.~\ref{fig:LIDC}(e) the \textit{curvature} is barely affected by the gate voltage, indicating that what is altered is just the prefactor $I_0$ and not the \textit{shape} of the CPR. In fact, the simplest interpretation of the impact of $|V_g|$ is that it changes the number of transverse channels $N$ that carry the supercurrent, while $\bar{\tau}$ stays constant. This alters the prefactor $I_0(V_g)$ which is given by
\begin{equation}
I_0(V_g)\ =\ \frac{e\Delta^{\ast}}{\hbar}\,N(V_g).
\label{eq:I0}
\end{equation}
Using Eqs.~\ref{eq:cprshortball} and~\ref{eq:I0}, we extract  $N(V_g)$ versus gate voltage from the data in Fig.~\ref{fig:LIDC}(e) and obtain $N(0)=187$ at $V_g=0$. 
%Also, we have determined the Fermi wavelength $\lambda_F$ by an independent Hall measurement on a different chip from the same wafer, with the Al film fully removed. If we assume that the electron density in the quantum well remains unchanged after stripping the Al film, the number of transverse channels deduced from $\lambda_F$ is roughly four times $N(0)$~\cite{Note1}. 
%This observation is consistent with the measured value of $I_0$, if the distribution of  channel transparencies is bimodal with 1/4 of highly and 3/4 of weakly transmissive channels~\cite{Note1}. The latter contribute little to both $I(\varphi)$ and $L$. Such a bimodal distribution of highly and weakly transmissive channels has been observed also in Refs.~\cite{Nichele2020,dartiailh2020missing}. Changing $V_g$ to -2.0\,V depletes the channel just by 25\%, because the junctions are effectively screened by the Al islands. At $V_g<-2.0\,$V the weakest junctions become resistive and spoil the quality factor.
This number is very close to the value $N=[(2e^2/h)R_{sh}]^{-1}=193$ obtained from the Sharvin resistance in the normal state, $R_{sh}=R_n=66.9$~$\Omega$. The good agreement between the two estimates of $N$ demonstrates that $I_0$ is not suppressed by environment effects as often observed in $I(V)$-characteristics. The normal state resistance allows us to estimate the product $I_cR_n$, which can be compared with the theoretical ballistic limit  $\pi\Delta^{\ast}/e$. We find that the measured  $I_cR_n=295$~\textmu V, which is 72\% of the ballistic limit for  $\Delta^{\ast}=130$~\textmu eV. This fraction is close to that (69\%) observed in Ref.~\cite{Mayer2020} for clean junctions with a length comparable to ours.

\begin{figure}[tb]
\includegraphics[width=\columnwidth]{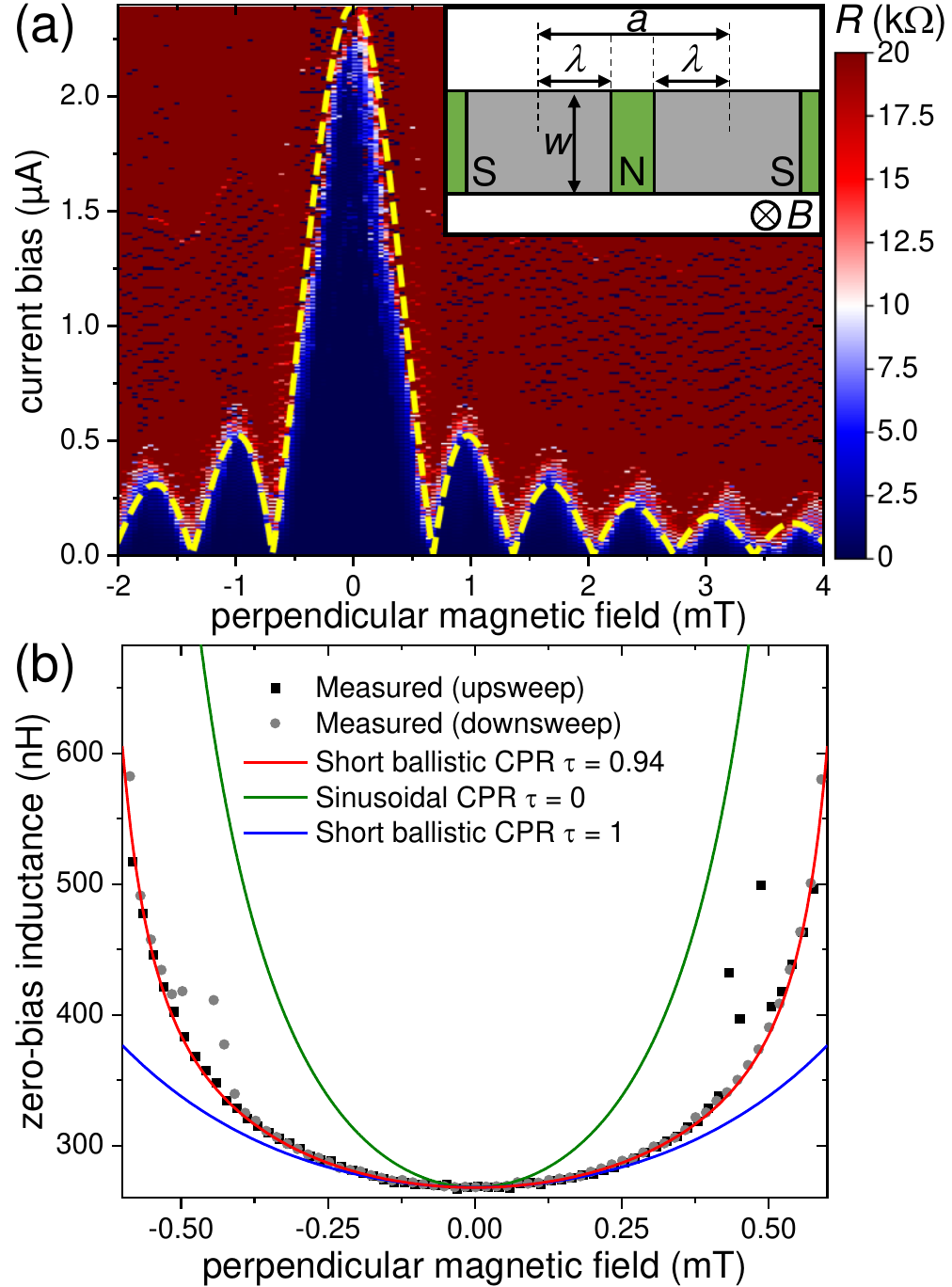}
\caption{(a) Color plot of the differential resistance plotted as a function of perpendicular magnetic field $B_{\perp}$ and current bias. The dashed yellow line shows the expected critical current  $I_c(B_{\perp})$ for a rectangular junction with effective length $a=960$~nm and width $w=3.15\,\mu$m, see inset.  (b) Zero-bias Josephson inductance $L$ as a function of $B_{\perp}$ for the central lobe in the diffraction pattern (symbols) together with the curves deduced from Eq.~\ref{eq:cprshortball} for $\bar{\tau}= 0.94$  (red), $\bar{\tau}\rightarrow 1$  (blue) and $\bar{\tau}\rightarrow 0$  (green). For the latter curve, the parameter $I_0$ has been rescaled by a factor 2.06 to match the measured zero-field inductance~\cite{Note1}.
}
\label{fig:Diffr}
\end{figure}

A hallmark of the Josephson effect and an important indicator for junction homogeneity is the modulation of the critical current $I_c(B_{\perp})$ by quantum interference in a perpendicular magnetic field $B_{\perp}$.  Figure~\ref{fig:Diffr}(a) shows the JJ array resistance measured in DC as a function of $B_{\perp}$ and  $I$ at $T=100$\,mK. The resistance is obtained by numerical differentiation of IV-characteristics.  The diffraction pattern $I_c(B_{\perp})$ is visible as the boundary between near-zero and finite resistance regions. It matches the Fraunhofer pattern well known from tunneling junctions: $I_c(B_{\perp})=I_c(0)|\sin(\pi\Phi/\Phi_0)/(\pi\Phi/\Phi_0)|$. 
This is not by accident:  the normalized diffraction pattern $I_c(B_{\perp})/I_c(0)$ calculated from Eq.~\ref{eq:cprshortball} by integrating the current density over the width of the junctions turns out to be independent of $\bar{\tau}$~\cite{Note1}. 

 The period of the diffraction pattern is determined  by the flux $\Phi=awB_{\perp}$ within the effective junction area, where $w$ and  $a$ are width and effective length, respectively. 
%The latter is given by the gap between the islands plus twice the effective penetration length $\lambda$, see inset Fig.~\ref{fig:Diffr}(a). 
From the lobe periodicity in Fig.~\ref{fig:Diffr}(a) we find  $a=960$~nm. This is close to the lattice period $a_0=1.1\,\mu$m of the array  %{\red footnote seems note to be compatible with cite!}
  \footnote{The length $a$ is usually estimated as the spacing between the Al islands plus twice the effective magnetic penetration depth. In large films of thickness $d$ the field penetrates up to a length~\cite{Pearl1964,Tinkhambook}  $\lambda_{\perp}=\lambda_L^2/d$, where $\lambda_L$ is the London penetration depth. In our case $\lambda_L=220$~nm, as obtained from independent measurements of the kinetic inductance of the same heterostructure,  which is also in reasonable agreement with the estimate from the normal state resistance.  Since $\lambda_{\perp} \simeq 8$~\textmu m is much larger than the Al island size, the field penetrates them almost completely. Thus, the value of the effective junction length $a$ is expected to be close to the junction periodicity, as verified by the Fraunhofer pattern measurement and confirmed by the good agreement between red curve and experimental points in Fig.~\ref{fig:Diffr}(b).}.
 At $B_{\perp}=0$, most JJs switch to normal resistance at current bias of  2.4~\textmu A, which is considerably less than than the critical current $I_c = 4.41$~\textmu A defined above.
%(for $I_0=5.88\,\mu$A and $\bar{\tau}=0.94$) determined near equilibrium from inductance measurements. 
The reason for the discrepancy is once again the presence of weaker junctions. Once they switch to normal resistance, their dissipation heats the remaining junctions, leading to a runaway process that rapidly brings the whole array into the normal state.

In order to substantiate our evaluation of the average transparency~$\bar{\tau}$ against yet another observable, we turn now to the dependence of the zero bias Josephson inductance on $B_\perp$.
In contrast to the critical current, the diffraction pattern in $L$ turns out to be very sensitive to $\bar{\tau}$. This is demonstrated in Fig.~\ref{fig:Diffr}(b), displaying the measured inductance $L(B_{\perp})$ for the central lobe (symbols) together with the expectation for $L(B_{\perp})$ (red solid line) using the CPR in Eq.~\ref{eq:cprshortball}. Without further adjustment of the previously determined parameters $I_0$, $\bar{\tau}$ and $a$, we find an excellent agreement that corroborates our analysis. The green and blue curve show instead the limiting cases of perfect opacity ($\bar{\tau}\rightarrow 0$, sinusoidal CPR, green curve) and perfect transparency ($\bar{\tau}\rightarrow 1$, blue curve) case, respectively. For the latter cases, the value of $I_0$ has been rescaled to obtain the measured value of zero-field inductance. It is clear that, even then, CPRs with values of $\bar{\tau}\neq 0.94$ cannot reproduce the experimental data. Instead,  Eq.~\ref{eq:cprshortball} with $\bar{\tau}=0.94$ correctly describes not only the equilibrium $L(B_{\perp})$, but also the curves for finite current bias~\cite{Note1}. 

In conclusion, we have shown that the Josephson inductance is a sensitive and versatile probe of the Andreev spectrum in short ballistic SNS junctions. The temperature, bias, gate voltage, and perpendicular field dependence of the Josephson inductance can be quantitatively described in terms of a short ballistic weak link with nearly perfect transmission. Inductance measurements enable the direct determination of the induced gap and of the junction transparency. Our experimental scheme can be easily combined with standard DC setups and allows for a simultaneous measurement of  DC transport properties.

\begin{acknowledgments}
We thank  Asbj\o rn Drachmann, Charles M. Marcus and Marco Aprili for stimulating discussions. Work at UR was funded by the Deutsche Forschungsgemeinschaft (DFG, German Research Foundation) – Project-ID 314695032 – SFB 1277 (Subproject B08). Work completed at Purdue University is supported by Microsoft Quantum.
\end{acknowledgments}

\bibliography{biblio}% Produces the bibliography via BibTeX.

%\pagebreak
\clearpage
\newpage

\onecolumngrid
\begin{center}
\textbf{\large Supplemental Material: Josephson inductance as a probe for highly ballistic semiconductor-superconductor weak links}
\end{center}

\twocolumngrid
\beginsupplement

\section{Wafer growth and characterization}

\begin{figure*}[tb]
\includegraphics[width=2\columnwidth]{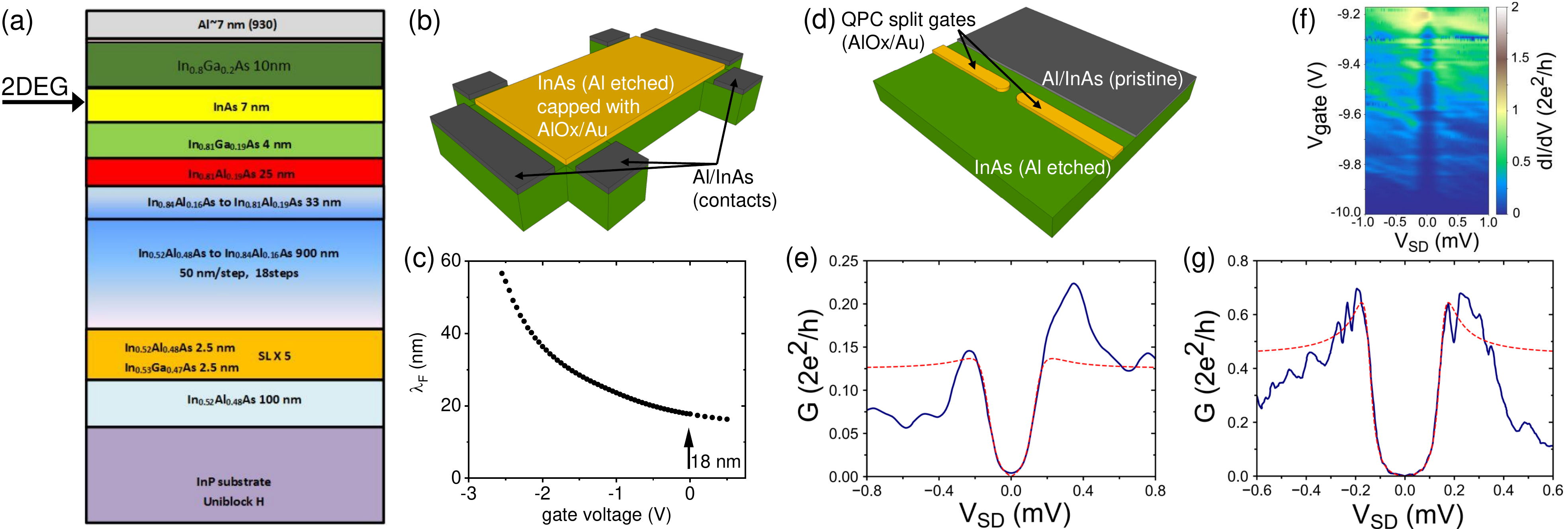}
\caption{(a) Growth sequence for the heterostructure under study, with the correct stoichiometry and layer thickness. The 2DEG is located near the InAs layer. (b) Sketch of the Hall bar used for Hall measurements. To characterize the 2DEG only, we etched out the epitaxial Al, everywhere but at the contacts.  (c) Gate voltage dependence of the Fermi wavelength, measured via a Hall measurement on a test Hall bar from the same wafer used to produce the Josephson junction arrays.  (d) Sketch of the device used for single channel spectroscopy. A global top gate is present but not shown for simplicity. A NS interface is obtained by etching out the epitaxial Al on half of the sample, exposing the 2DEG. A few hundreds nanometers from the interface a quantum point contact (QPC) is fabricated by patterning of a AlOx/Al film. (e,g) Two differential conductance curves (solid blue lines) measured as a function of voltage bias on a QPC kept at $T=100$~mK and $V_g=-4$~V (panel (e)) and $T=40$~mK and $V_g=-5.5$~V (panel (g)). The red dashed curves are fits with the Dynes formula. The fit parameters are $\Delta^{\ast}=137$~\textmu eV and $\Gamma=60$~\textmu eV (panel (e)) and $\Delta^{\ast}=150$~\textmu eV and $\Gamma=28$~\textmu eV (panel (g)).  (f) The color plot shows the differential conductance  as a function of the gate voltage and voltage bias, measured at 300~mK on a different QPC sample (but from the same wafer).  A well-defined superconducting gap is observed in the opaque tunnel regime, for gate voltage below -9.25~V.
}
\label{fig:stack_lambdaF}
\end{figure*}

The layer sequence of the hybrid heterostructure under study is depicted in Fig.~\ref{fig:stack_lambdaF}(a). Between the top 10~nm-thick In$_{0.8}$Ga$_{0.2}$As barrier and the  \textit{in situ}-grown epitaxial Al layer there are two monolayers of GaAs. 

The quantum well was studied in a top-gated hall bar geometry (sketched in Fig.~\ref{fig:stack_lambdaF}(b)), on a test sample fabricated from the same wafer. Here the Al film was etched selectively and the laid open area was covered with a 40~nm aluminum oxide layer and a Au film as top-gate. A maximum mobility of 22000~cm $^2$/Vs is observed at density $n = 0.5\cdot 10^{12}$~cm$^{-2}$ for gate voltage $V_{g}=-1.8$~V, resulting in a mean free path length $\ell_e \approx 270$~nm. The gate dependence of the electron density allows us to deduce that of the Fermi wavelength, shown in Fig.~\ref{fig:stack_lambdaF}(c).

Following Ref.~\cite{Kjaergaard2016}, we characterized the proximity-induced gap and the Andreev reflection between a few-channel NS junction. To this end we fabricated a quantum point contact near a NS interface, which was obtained simply by etching out the epitaxial Al in half of a Hall bar.  A sketch of the device is shown in Fig.~\ref{fig:stack_lambdaF}(d). Notice that a global top gate is present but not shown in the Figure for ease of readability.  Figure~\ref{fig:stack_lambdaF}(f) shows a color plot of the differential conductance versus gate voltage and bias voltage, measured at 350~mK. In the opaque regime, for gate voltages below -9.25~V, the differential conductance is proportional to the density of states: we deduce therefore a gap of approximately 140~\textmu  eV, as deduce by fitting several curves. The uncertainty of this induced-gap determination is at least 20~\textmu  eV. As an example, Fig.~\ref{fig:stack_lambdaF}(e,g) shows two of such spectra with the corresponding fits. 

\section{Sample fabrication}

The Josephson junction array was fabricated by defining a 3.15~\textmu  m-wide mesa by electron beam lithography and a standard (orthophosphoric acid~:~citric acid~:~hydrogen peroxide~:~distilled water~=~1.2~:~22~:~2~:~88) wet etching solution. Such etching step completely removes the 2DEG outside the mesa. In the following step, the gaps between Al islands have been selectively etched using the etchant type D from Transene Company. The entire array was covered with a 40~nm aluminum oxide layer via atomic layer deposition and a 5~nm Ti/120~nm Au metal layer operating as a top-gate.

\section{RLC circuit design}

To detect small inductances at low temperature, we have designed a circuit based on a cold RLC resonator, integrated directly on the sample holder and connected to a cold ground, which allows us to extract the sample inductance from the resonance frequency. This scheme is an adaptation of the setup in Ref.~\cite{Meservey1969}, here implemented using digital lock-in amplifiers, which allow us to obtain the full resonance spectrum, and thus to accurately determine not only the center frequency (and thus the inductance), but also the quality factor, from which we deduce the effective resistance. Importantly, in the range of parameters chosen for our RLC circuit ($L_0=382$~nH, $C_0=4$~nF, $R_{Dj}=1$~k\textOmega~for $j=1,2,3,4$), the total series resistance of the  tank circuit ($R+R_0$ in Fig.~1(e) of the main text, with $R_0$ being the low temperature resistance of the external coil) must be kept below few ohms in order to obtain a sufficiently high $Q$-factor. This is four orders of magnitude less than the normal resistance $R_N=157$~k$\Omega$ of the array. Therefore, all the inductance measurements here reported are performed in a nearly perfectly dissipationless transport regime.  

%Our RLC circuit is sketched in Fig.~\ref{fig:firstfig}(e). A home-made Cu coil with inductance $L_{0}$ and a capacitor with capacitance  $C_{0}$ define the working point in frequency, namely $f_0=1/(2\pi\sqrt{L_{0}C_{0}})$, which is approximately 4~MHz~\cite{Note1}. The Josephson inductance in series to $L_{0}$ produces a  downshift in the center frequency of the resonance spectrum. The inductance $L_0=382\,$nH  in series to the array has been measured separately with the array replaced by an Al bond wire~\cite{Note1} and has been subtracted from the measured inductance to obtain the data reported in what follows.

%The RLC resonator is integrated directly on the sample holder and connected to a cold ground. The input resistors $R_{D1}$--$R_{D4}$ decouple the resonator from the cryogenic transport lines (dashed lines in Fig.~\ref{fig:firstfig}(e)). In this way, the center frequency and width of the resonance peak in the spectrum only weakly depend on the relatively large stray reactance of the cryogenic wiring.  The capacitor $C_0$ also isolates the cold ground from the source contacts at low frequencies, allowing for DC measurements. The resistance of the resonating circuit in series to the JJ array is about $R_0\simeq0.3\,\Omega$. 

\begin{figure}[tb]
\includegraphics[width=1\columnwidth]{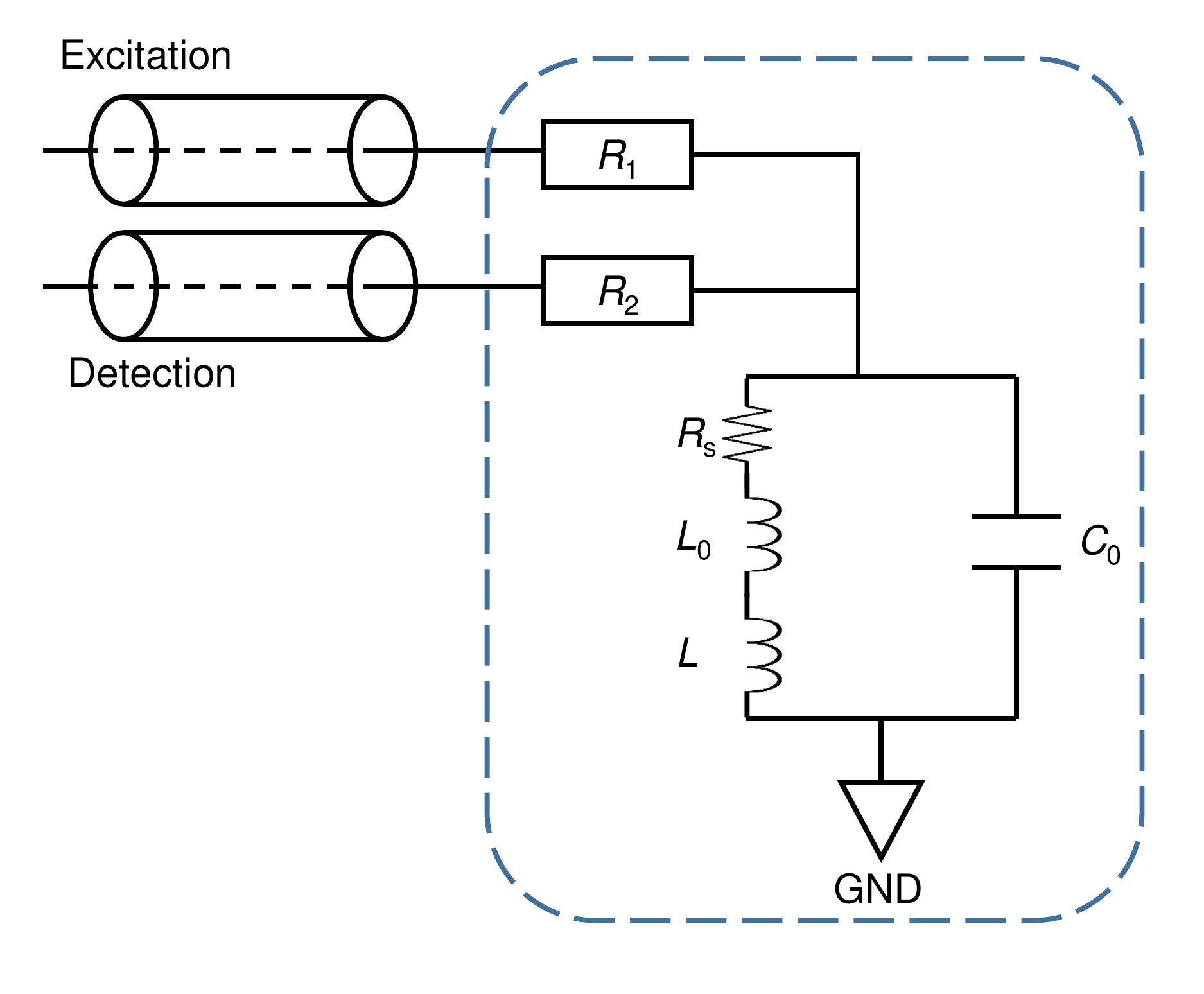}
\caption{Sketch of the model circuit. The coax wires represent the cryostat cables.  The part within the dashed blue line is the cold section of the resonant circuit. The total inductance $L_T$ is given by the sum of coil inductance $L_0$ and the sample kinetic inductance $L$. The capacitance $C_0$ is chosen in order to produce the highest resonance frequency compatible with the available electronics ($\approx 4$~MHz in our case).  The resistors $R_1$ and $R_2$ decouple the resonator from the rest of the cryostat. If their parallel resistance is much bigger that the resonance impedance $Z_m$ of the resonator, then the resonator can be considered in good approximation a series RLC, with resonance frequency $f_0=(2\pi\sqrt{L_TC_0})^{-1}$, $Q$-factor $Q=R_s^{-1}\sqrt{L_T/C_0}$, and resonance impedance $Z_m=R_sQ^2$. In order to observe a well defined resonance peak, we seek for a $Q$-factor bigger than one. This limits the maximum tolerable $R_s$.
}
\label{fig:ideal}
\end{figure}

The choice of the RLC circuit parameters needs to keep into account the expected inductance to be measured, the total resistance and the available frequency range. Figure~\ref{fig:ideal} shows the idealized circuit, where for simplicity we ignore the voltage probe lines. The cold part of the circuit is highlighted with a dashed blue line. The resistors $R_1$, $R_2$ (playing the role of $R_{D1}$ and $R_{D2}$ in Fig.~1(e) of the main text) are used to decouple the resonator from the cryostat cables. In this way, the center frequency and width of the resonance peak in the spectrum only weakly depend on the relatively large stray reactance of the cryogenic wiring. As long as  the resistances $R_1$, $R_2$ are sufficiently high (that is, higher than the resonance impedance of the tank, see below), the resonator shall behave as a \textit{series} RLC circuit, since in this case the only relevant  resistance $R_s$ is placed in series to both inductance and capacitance. $R_s$, in series to the resonator, includes the sample resistance $R$ plus the resistance $R_0$ of the entire resonating circuit. This latter is the sum of coil resistance, resistance of bonding wires and contact resistance between  bonding wire and epitaxial aluminum. Apart from the sample contribution, in our case $R_s\approx 0.3$~$\Omega$.

The capacitor $C_0$ also isolates the cold ground from the source contacts at low frequencies, allowing for DC measurements. The inductance $L_0$ is obtained from a home-made Cu coil.
The total inductance $L_T$ is given by the sum of an external inductor $L_0$ plus the interesting Josephson inductance $L$. The product of typical $L_T$ and capacitance $C_0$ is chosen in such a way to keep the center frequency $f \equiv 1/2\pi\sqrt{L_T C_0}$ of the resonance peak of the order of 4~MHz, that is, compatible with the range of our electronics (Zurich MFLI lock-in, with maximum frequency of 5~MHz). 

A parameter of interest is the sensitivity $S$, which we define as $$S \equiv \frac{\delta f}{\Delta f},$$ where $\delta f$ is the tiny frequency shift produced an inductance change $\delta L$, and $\Delta f$ is the width of the resonance peak. For a \textit{series} RLC circuit the $Q$-factor is 
\begin{equation}
Q=\frac{1}{R_s}\sqrt{\frac{L_T}{C_0}}.
\label{eq:Qfact}
\end{equation}
Therefore the sensitivity is 
\begin{equation}
S=\frac{\partial f}{\partial L}\delta L \frac{1}{\Delta f}=\frac{\pi f_0}{R_s} \delta L,
\label{eq:sens}
\end{equation}
where it is clear that, once $f_0$ is given, the sensitivity will only depend on $R_s$. This is valid as long as Eq.~\ref{eq:Qfact} is valid, i.e., in the limit that the parallel resistance of the input resistors $R_1$ and $R_2$ (plus those in front of the voltage probe lines if present) is much bigger than the peak impedance of the tank, see below.
%If we assume $f_0=2.5$~MHz, $R_s=0.1$~$\Omega$ and $\delta L=100$~pH, the sensitivity is of the order of 7$\cdot 10^{-3}$. This means that the shift of the lorentzian due to an addition of 100~pH is nearly 1\% of its width, which should be easily detectable unless the signal is too noisy.

In order to observe a resonance peak in the first place, the resonator must be underdamped, that is, $Q\gg 1$. This means that $L_T\gg R_s/\omega_0$. This sets a lower limit for $L_0$. Increasing arbitrarily $L_0$, however, does not help improving $S$, since this latter is independent on the inductance once $f_0$ is fixed. Also, an increase in $L_0$ is in general accompanied by an increase in $R_s$, since most of the resistance comes from the inductor. The design of the inductor has therefore to produce a sizable inductance with the smallest possible resistance and without a ferromagnetic core that would be incompatible with measurements in magnetic field. Therefore, we used a thick home-made coil starting from a pure Cu wire.

Let us now consider the condition for the external resistors $R_1$ and $R_2$. They will effectively decouple the resonator from the external cables if the maximum impedance $Z_m$ of the resonator (e.g. at the resonance) is much smaller than the parallel $R_p$ of $R_1$ and $R_2$ (and eventually the resistors in front of the voltage probe lines). In this limit Eq.~\ref{eq:Qfact} is valid and the circuit behaves as a series RLC circuit. For this kind of circuit the maximum tank impedance is 
\begin{equation}
Z_m=R_sQ^2=\frac{L}{R_sC_0}=\frac{4\pi^2f^2L^2}{R_s}.
\label{eq:zm}
\end{equation} 
Assuming $L_T\approx 600$~nH, $R_s=0.3$~$\Omega$ and $f_0=3$~MHz we obtain at the resonance that $Z_m\approx 400$~$\Omega$. For this reason, the decoupling resistors have been chosen to have 1~k$\Omega$ resistance.  

\section{Dissipation and $Q$ factor}
\begin{figure}[tb]
\includegraphics[width=1\columnwidth]{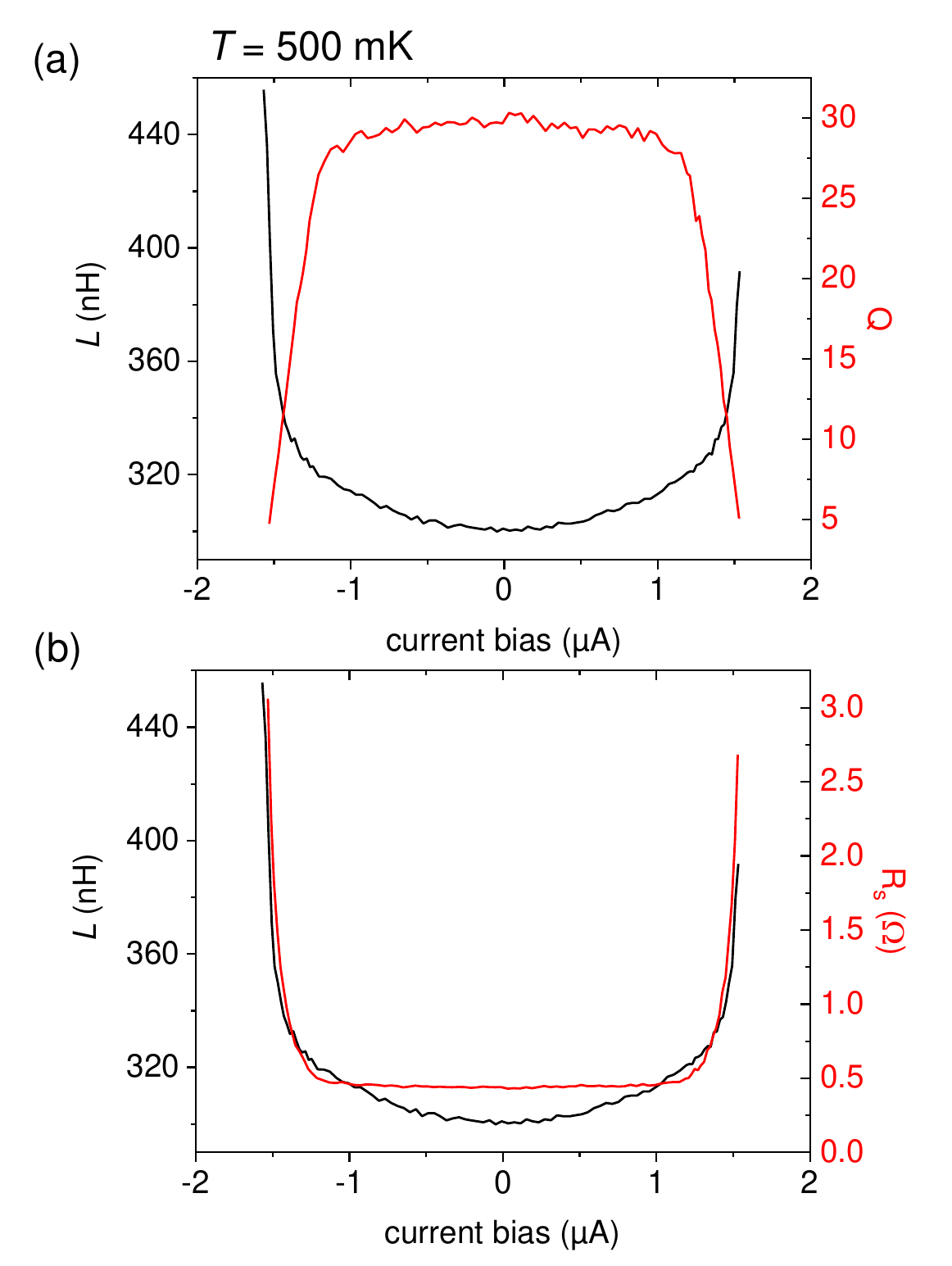}
\caption{(a) The graph shows the finite-bias Josephson inductance $L(I)$ (obtained from the center frequency of the resonance spectrum) together with the measured $Q$ factor (obtained from the width of the resonance peak). Notice that at low bias the $Q$ factors saturates to a maximum value of nearly 30. (b) Circuit resistance $R_s$ calculated from the $Q$ factor in the limit of a RLC circuit in series. As explained in the text, such limit is not strictly valid for low sample dissipation, therefore the saturation value at low current bias slightly overestimates $R_s$.}
\label{fig:RsandQ}
\end{figure}

As mentioned in the main text, the resonant spectrum of the RLC circuit provides not only the sample inductance (from the center frequency of the resonance peak) but also the  circuit resistance  (from the $Q$ factor). An an example, Fig.~\ref{fig:RsandQ} shows the $Q$ factor (panel (a)) and the resistance  $R_s$ (panel (b)) of the resonant circuit, measured together with the finite-bias Josephson inductance at $T=500$~mK. In this case, $Q$ is computed directly from the resonance width, while $R_s$ is assumed to be $R_s=Q^{-1}\sqrt{(L+L_0)/C_0}$, where $L_0$ is the external inductance in series, $L$ is the Josephson inductance and $C_0$ is the capacitance of the tank. An upper limit to the $Q$ factor is given by ($i$) the residual resistance of the RLC tank (mostly due to the coil resistance) and ($ii$) the fact that the circuit can be approximated as a series RLC as long as $Z_m$ is much smaller that the parallel of all the resistors $R_{D1}$--$R_{D2}$. In the case of Fig.~\ref{fig:RsandQ}, for low current bias (and thus very low $R_s$) the parallel resistance of four 1~k$\Omega$ resistors (250~$\Omega$) is comparable or even lower than $Z_m$. This means that in this regime, a further reduction of the resistance in series to the sample would not improve the $Q$ factor. In this sense, our system is close to its optimum. Indeed, the choice of the resistors in series is made precisely with this aim: once $R_s$ has been reduced as much as possible, the parallel resistance of the decoupling resistors must be of the order of the maximum $Z_m$. If they are larger, they would reduce the input signal without gain in the $Q$ factor; if they are smaller, this will suppress the $Q$ factor. 
%Rs_and_Q

\section{Determination of the external inductance $L_0$}
As discussed in the main text, it is important to experimentally determine the external inductance (mostly given by the home-made copper coil) $L_0$ with a calibration measurement. This calibration was performed just before the sample cool-down. To this end, we used the very same circuit, with an identical chip-carrier, whose source and drain pins were shorted with a bonding wire. From the center frequency of the resonance peak we deduced $L_0=382$~nH. Therefore, throughout this work the Josephson inductance $L$ is taken as the difference between the measured total inductance $L_{T}$(deduced from the center frequency of the resonance curve via automated fit) and $L_0$. We quantified possible sources of residual discrepancies between ($L_{T}-L_0$) and $L$: it turns out that the main discrepancy is due to the kinetic inductance of the epitaxial Al leads. For our Al-film thickness and lead geometry we estimate that such kinetic inductance is of the order of few~nH which is compatible with the scatter of the experimental points.

\section{Experimental uncertainty} 
\begin{figure}[tb]
\includegraphics[width=\columnwidth]{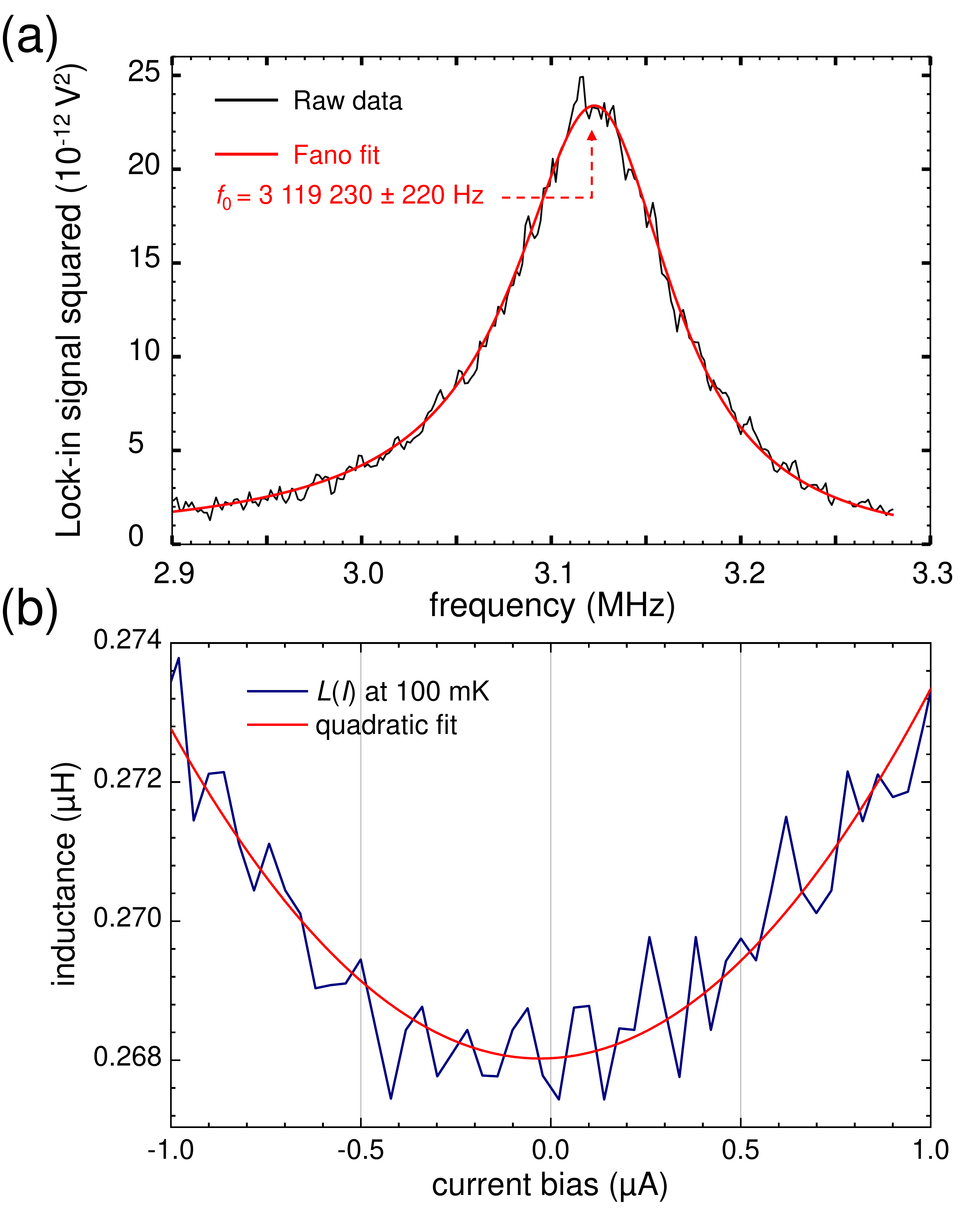}
\caption{(a) Exemplary spectrum (amplitude squared of the voltage signal) as measured from the lock-in (black curve), together with a Fano fit (red curve). Compared to a lorentzian fit, a Fano fit better accounts for the slight asymmetry of the resonance curve, which is due to the details of the real circuit. (b) Zoom-in on the $L(I)$ curve in Fig.~2 of the main text, together with a parabolic fit.}
\label{fig:uncer}
\end{figure}

The measurement of the inductance relies on the determination of the center frequency of the resonance spectrum for the RLC circuit. As long as the noise level is sufficiently low, the inductance uncertainty does not depend on it. As shown in Fig.~\ref{fig:uncer}(a), the fitting routine can determine the center frequency with an error of 220~Hz over 3 MHz, which translates into an inductance uncertainty of 0.1~nH.  If we look at a series of points in a bias sweep (as, e.g., the $L(I)$ curve at 100~mK in Fig.~2(a), see zoom-in graph in Fig.~\ref{fig:uncer}(b)) we notice that the scatter of the inductance values is of the order of 0.6~nH (taken as the standard deviation from the parabolic fit in Fig.~\ref{fig:uncer}(b)). This latter value keeps into account all fluctuations in the setup (DC bias, temperature, etc.) and should be considered as the closest measure of the uncertainty in the inductance.

\section{$IV$-Characteristics}
\begin{figure}[tb]
\includegraphics[width=\columnwidth]{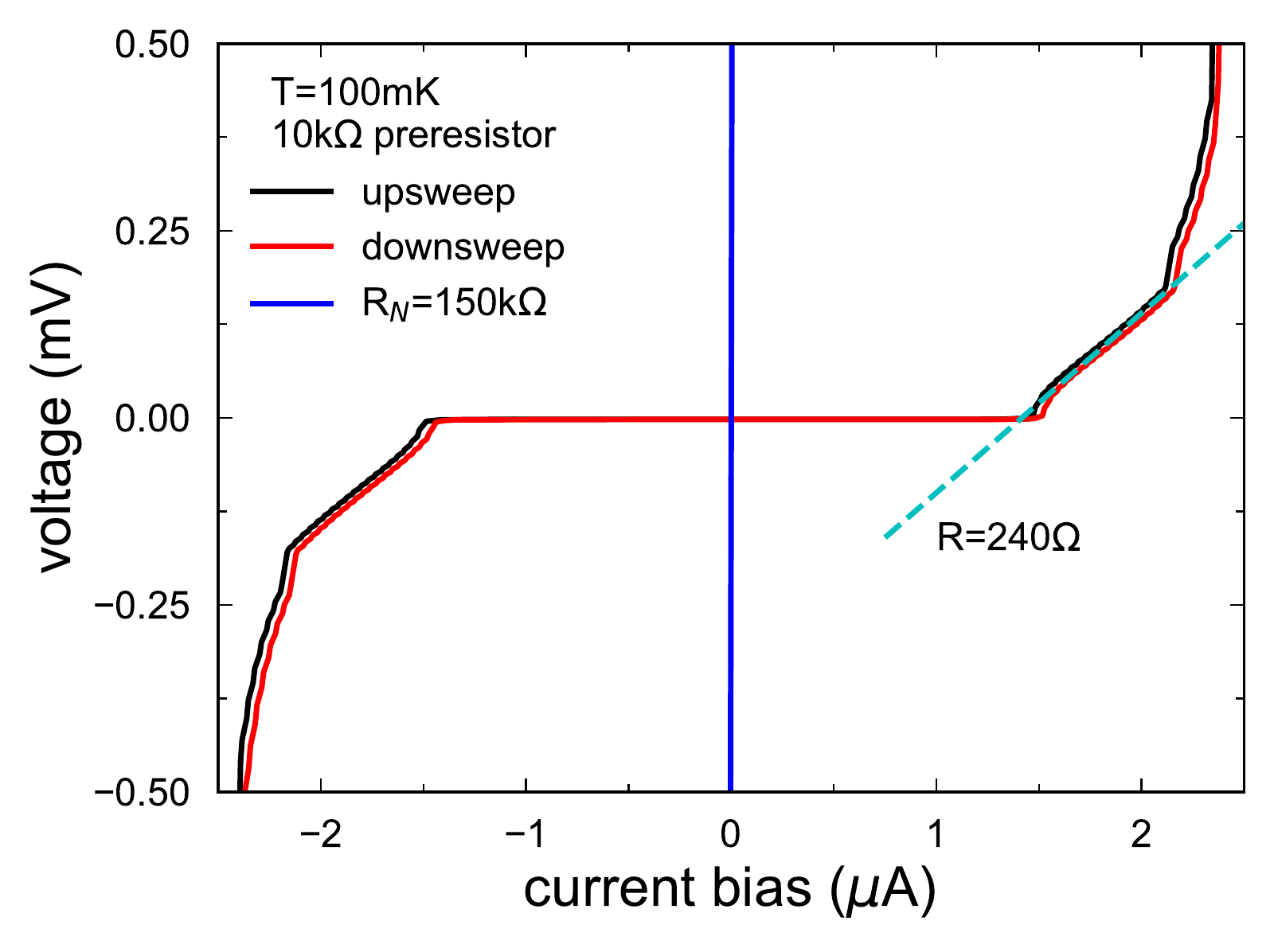}
\caption{$IV$-characteristics measured at $T=100$~mK in DC for current swept from negative to positive bias (black curve) and vice versa (red curve). For comparison we plot $V=R_NI$ with $R_N=157$~k\textOmega .}
\label{fig:ivchar}
\end{figure}

Figure~\ref{fig:ivchar} shows representative IV-characteristics of the array under study. The two curves refer to the current bias swept from negative to positive values (black curve, upsweep) and from positive to negative values (red curve, downsweep). The measurement is performed at a temperature of 100~mK. A current bias is obtained applying a voltage with a DC source via a 10 k$\Omega$ preresistor.

The slope of the IV-characteristics does not saturate to the normal resistance within the bias applied (see blue curve in Fig.~\ref{fig:ivchar}). This one has been limited to few microamperes in order to avoid accessive heating via the decoupling resistors $R_{1-4}$. The normal resistance has been measured at zero bias by temperature sweeps reported below. We notice a moderate hysteresis, which we attribute to heating effects and not to retrapping.  Josephson junctions based on epitaxial Al-InAs have by construction a small capacitance, therefore they are always in the overdamped regime. 

\section{Perpendicular magnetic field dependence}

\begin{figure}[tb]
\includegraphics[width=\columnwidth]{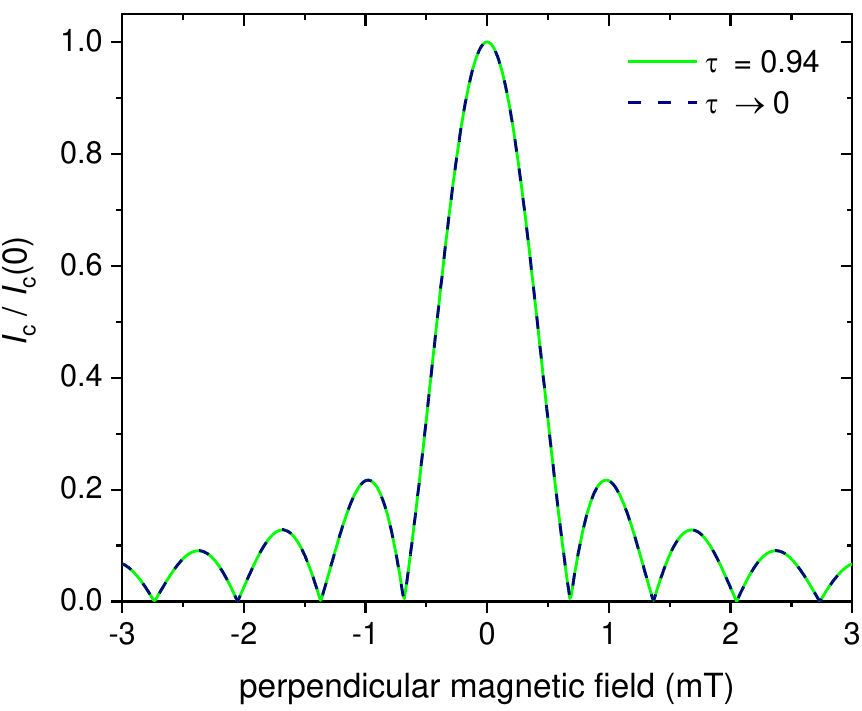}
\caption{Perpendicular magnetic field dependence of the maximum $I_c$ of the current-phase relation (CPR) normalized to its zero-field value for (dashed black curve) the CPR of Eq.~3 of the main text with $\bar{\tau}=0.94$ and for (green curve) the tunnel limit $I_c(B_{\perp})/I_c(0)=|\sin(\pi\Phi/\Phi_0)/(\pi\Phi/\Phi_0)|$, corresponding to $\bar{\tau}\rightarrow 0$ and $T \rightarrow 0$. The junction is assumed to be rectangular with length $a=960$~nm and $w=3.15$~\textmu  m. The two curves are hardly discernible, and this holds true for any value of $\bar{\tau}$.}
\label{fig:IcFP}
\end{figure}

\begin{figure*}[tb]
\includegraphics[width=2\columnwidth]{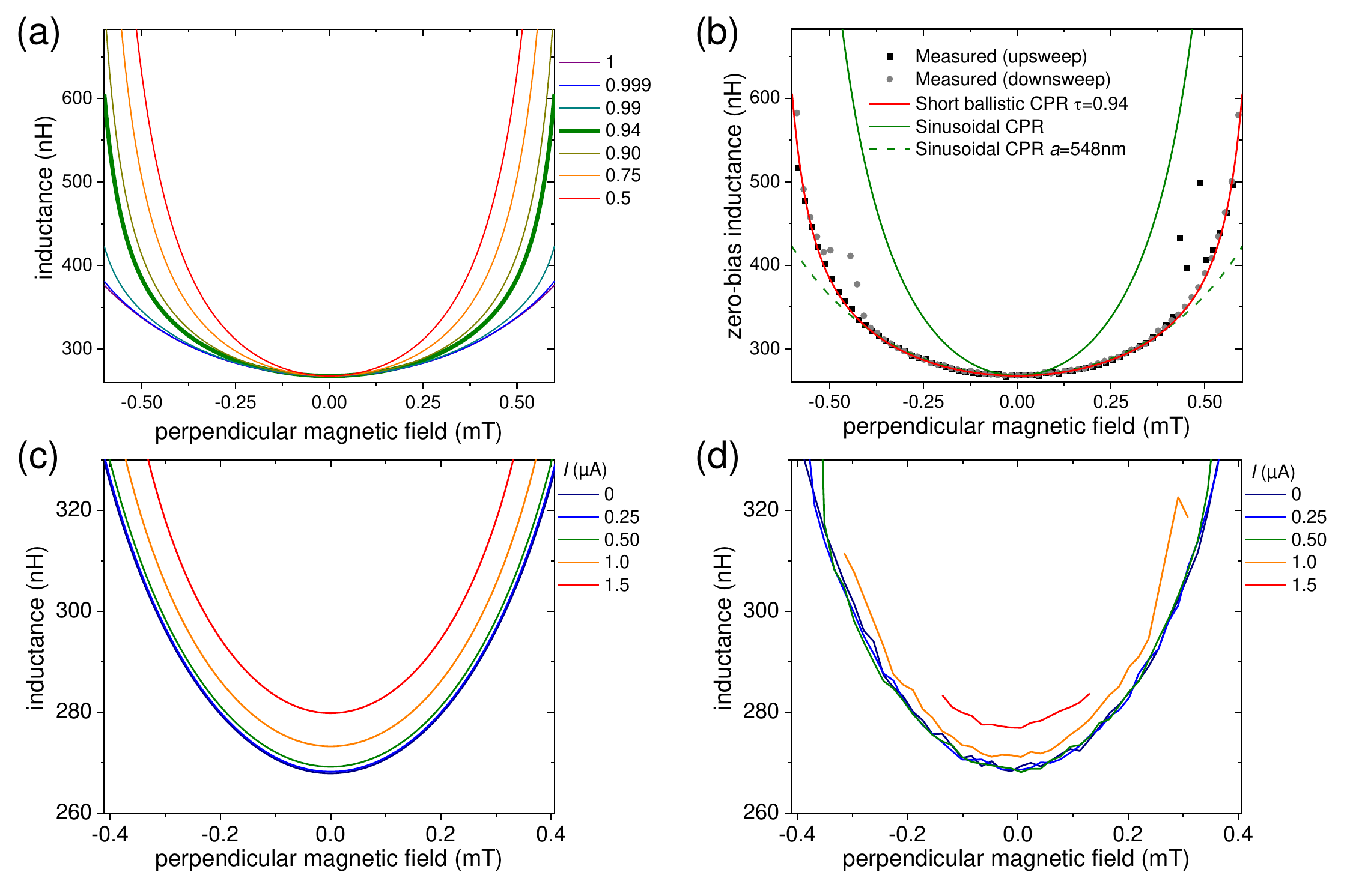}
\caption{(a) Perpendicular field dependence of the zero-bias Josephson inductance of 2250 junctions with the CPR described by Eq.~3 of the main text, computed in the limit $T\rightarrow 0$. Each curve corresponds to a different transparency coefficient $\bar{\tau}$. For ease of comparison, the curves have been rescaled to match the zero-field value of the $\bar{\tau}=0.94$ curve (this is equivalent to assume a different $I_0$). Notice the marked dependence on $\bar{\tau}$. (b) This panel reproduce Fig.~3(b) of the main text, with the addition (dashed green line) of a curve showing a fit of the experimental data using a sinusoidal CPR with the junction length $a$ as a free fitting parameter (best fit for $a=548$~nm). It is evident that not only the resulting $a$ does not match the lobe periodicity of the DC transport diffraction pattern (Fig.3(a) of the main text), but also that the curve cannot capture all the experimental points. (c) As in (a) but now $\bar{\tau}=0.94$ while the current bias takes finite values. (d) Corresponding experimental data measured in our array at $T=100$~mK.}
\label{fig:LvsBperp}
\end{figure*}

A peculiar feature of the current-phase relation (CPR) of short ballistic junctions is that the maximum supercurrent $I_c$ as a function of the perpendicular field $B_{\perp}$ is almost perfectly independent on the shape of the CPR, expressed, e.g.,~by the transparency $\bar{\tau}$. As a reference, Fig.~\ref{fig:IcFP} shows as a dashed-black curve $I_c(B_{\perp})/I_c(0)$ for the CPR considered in the main text (Eq.~3 with $\bar{\tau}=0.94$, $T\rightarrow 0$) with junction size $w=$3150~nm~$\times$~$a=960$~nm. As a comparison, the green curve corresponds to the tunnel limit for the same junction size ($\bar{\tau}\rightarrow 0$, $I_c(B_{\perp})/I_c(0)=|\sin(\pi\Phi/\Phi_0)/(\pi\Phi/\Phi_0)|$). The two curves are indiscernible, and this holds for any value of $\bar{\tau}$.

On the other hand, the Josephson inductance $L$ does depend on the CPR \textit{shape}, being $L$ proportional to the derivative of the inverse CPR. In order to calculate $L(B_{\perp})$ we start from the CPR in Eq.~3, $I=I_0f(\varphi)$. As in the textbook case of a rectangular junction in perpendicular field, we consider a local phase difference
\begin{equation}
\varphi(x) = \gamma +\left( \frac{2\pi aB_{\perp}}{\Phi_0} \right)x,
\label{eq:phiscale}
\end{equation}
where $x$ is the position along the junction width, $\gamma$ is the gauge-invariant phase difference between the superconducting leads and the linear term comes from the vector potential of a constant perpendicular field.
The current is obtained by integrating the CPR over the junction width 
\begin{equation}
I =\int_0^w (I_0/w)f(\varphi(x))dx \equiv I_0 g(\gamma,B_{\perp}),
\label{eq:int}
\end{equation}
where in the last step we defined the function $g(\gamma,B_{\perp})$, which is the average of $f$ over $[0,w]$ (we are assuming here that the junction is homogeneous).
The bias dependence of $\gamma=\gamma(I)$ is found by inverting the CPR in Eq.~\ref{eq:int} at a given $B_{\perp}$. 
The value of the inductance is therefore
\begin{equation}
L(B_{\perp})=\frac{\hbar}{2eI_0\left(\frac{\partial g}{\partial \gamma} \right)_{\gamma=\gamma(I)}},
\label{eq:Lind}
\end{equation}
where the dependence on $B_{\perp}$ is implicit in $g$ and $\gamma$.

Figure~\ref{fig:LvsBperp}(a) shows $L(B_{\perp})$ computed for several values of $\bar{\tau}$. Unlike the critical current $I_c$, the Josephson inductance diffraction pattern clearly depends on $\bar{\tau}$, in particular close to the perfect transparency.

Figure~\ref{fig:LvsBperp}(b) reproduces the curves of Fig.~3(b) of the main text, with an important addition. Here, we try to fit the experimental $L(B_{\perp})$ curve with a sinusoidal CPR, leaving $a$ as a \textit{free} parameter (green-dashed line). It is clear that, besides producing a $a$ value incompatible with the DC Fraunhofer pattern ($a=548$~nm), the sinusoidal CPR cannot suitably reproduce the experimental data.

Finally, the CPR in Eq.~3 correctly reproduces not only the zero-bias   $L(B_{\perp})$ curve, but also the same curve at finite current bias. This is shown in Fig.~\ref{fig:LvsBperp}(c) (computed curves) and Fig.~\ref{fig:LvsBperp}(d) (experiment). In particular, it is correctly reproduced the very weak dependence on the bias for small and moderate current bias.

%Subsequent theoretical developments predicted topological transitions also in engineered semiconducting/$s$-wave superconducting hybrid systems where time-reversal symmetry is broken in the presence of spin-orbit interaction (SOI) and a finite magnetic field~\cite{FuKane2008}.

\section{Impact of weak junctions on the Fraunhofer pattern}
\begin{figure}[tb]
\includegraphics[width=1\columnwidth]{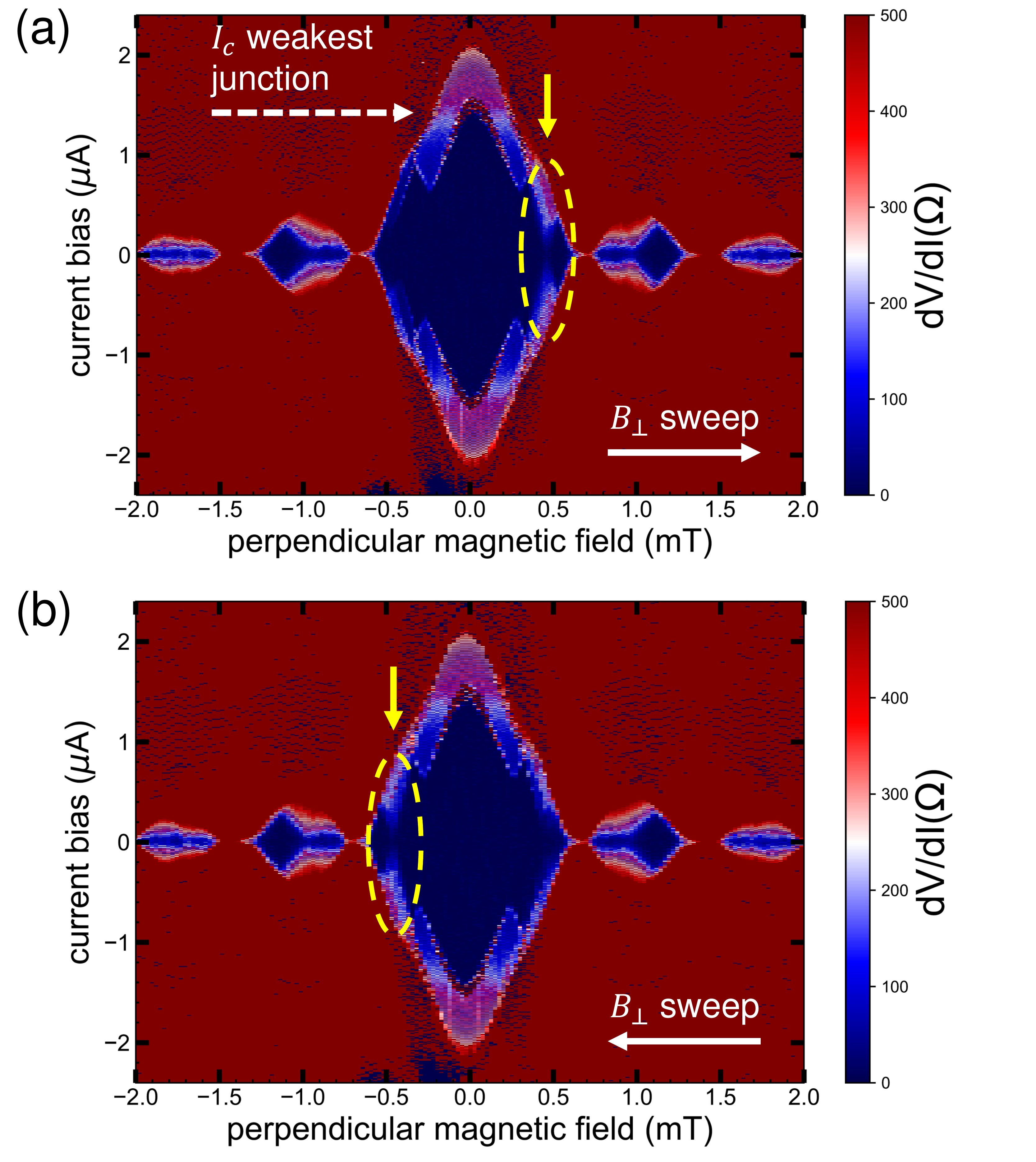}
\caption{(a) The color plot reproduces Fig.~3(a) of the main text, but with a reduced resistance rage. The resulting exaggerated color contrast reveals the resistance steps produced be one or few weaker junctions. (b) The same measurement, repeated in backsweep ($B_\perp$ swept from positive to negative values). We notice that at least one dissipative feature (marked with a vertical yellow arrow) within the main lobe is hysteretic in magnetic field.
}
\label{fig:hc}
\end{figure}

Figure~\ref{fig:hc}(a) shows the same graph as in Fig.~3(a) of the main text, but with an exaggerated color contrast to show the small resistance contribution of the first weak junction as a function of current bias and perpendicular magnetic field. Here the current is swept from negative to positive values (bottom to top in the plot) as well as the magnetic field (left to right). Figure~\ref{fig:hc}(b) refers to the backsweep in magnetic field (right to left). First of all, we notice that there are dissipative features appearing at currents below the main transition (here the boundary with the red region of the plot). This indicates that one or few junctions have a reduced critical current. The resistance step is in fact compatible with one or two junctions, whose individual normal resistance is about 66~$\Omega$. The resistance steps have a complex dependence on the magnetic field, producing the structures within the main lobe. One of these structures (indicated with a vertical yellow arrow in the graph) extends almost down to zero bias. Interestingly, it shows an hysteretic behavior in magnetic field, since it is observed for the opposite sign of $B_{\perp}$ in the backsweep of Fig.~\ref{fig:hc}(b). For this reason, we argue that these complex features are due to the penetration of a vortex in one island, which may affect the field in the junction and its transmission. The thorough understanding of such features goes beyond the scope of this work, since they are not relevant for our discussion. We just remark that it is such feature that is responsible of the few points (near $\pm 40$~mT) in Fig.3(b) where the two $B_{\perp}$ sweep direction do not match.

\section{Temperature dependence of the induced superconducting gap}

\begin{figure}[tb]
\includegraphics[width=1\columnwidth]{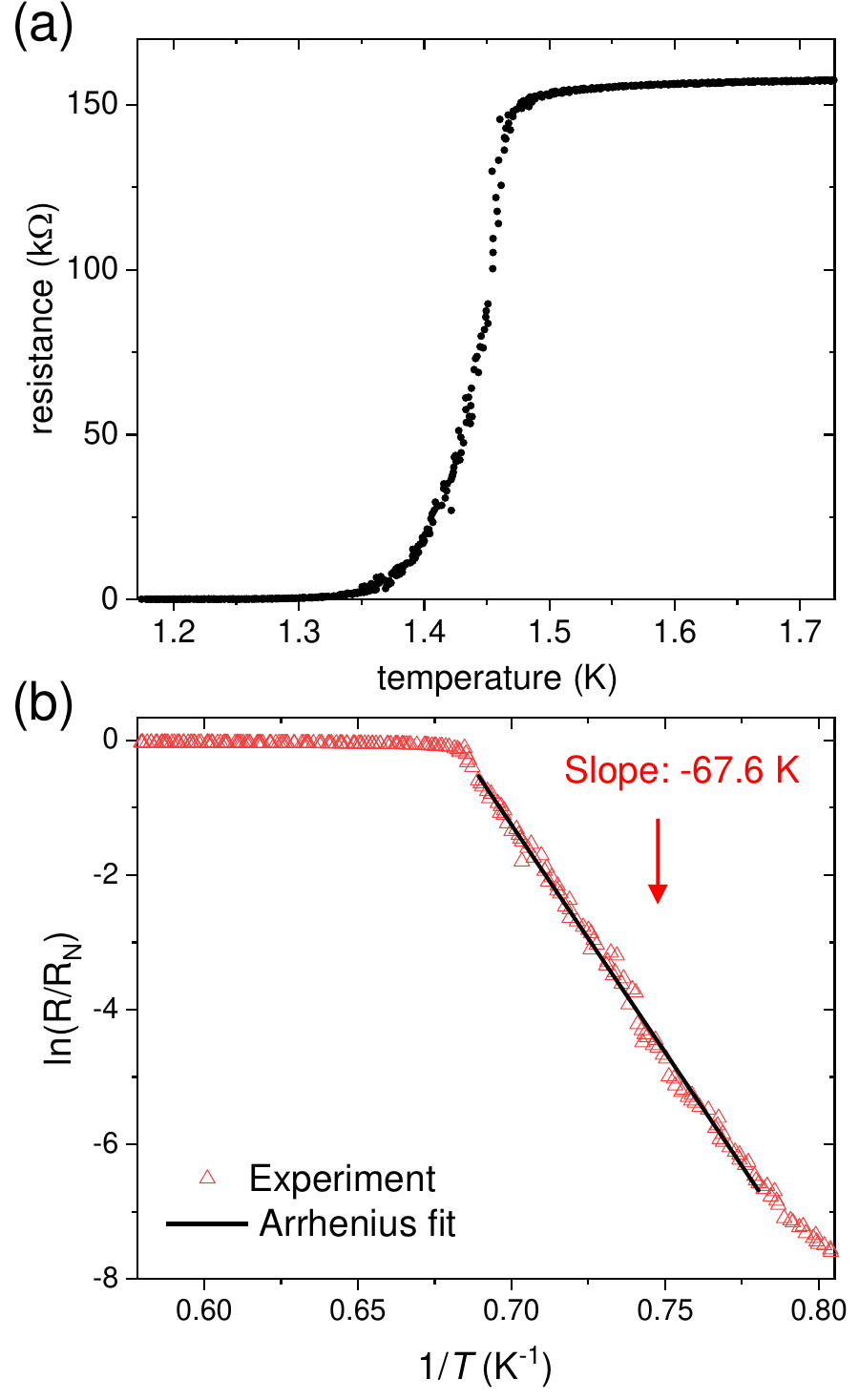}
\caption{(a) Resistance as a function of temperature for the full array. The total resistance in the normal state is 157~k$\Omega$, corresponding to a normal resistance per junction of about 66.9~$\Omega$ (after subtracting the contribution of epitaxial Al/InAs islands, which add an additional resistance of 6.5~k$\Omega$).  The temperature corresponding to half of the normal resistance $T_{0.5R_N}=1.44$~K is taken as critical temperature (b) Arrhenius plot near the foot of the transition. The experimental points are nicely aligned, showing an activation energy of about $68$~K$\times k_B$. The deviation at low temperatures (bottom right) can be simply attributed to an offset of few ohms (the plotted data are as measured).
}
\label{fig:roft}
\end{figure}

The calculation of the temperature dependence of $L(0)$ requires that of the induced gap $\Delta^{\ast}$. This latter does not follow the dependence expected from the BCS theory.  $\Delta^{\ast}(T)$ and $\Delta_{\text{Al}}(T)$ are related via~\cite{Chrestin1997,Aminov1996,Schaepersbook,KjaergaardPRAPPL17}
\begin{equation}
\Delta^{\ast}(T)\approx \frac{\Delta_{\text{Al}}(T)}{1+\gamma_B\sqrt{\Delta^2_{\text{Al}}(T) -\Delta^{\ast 2}(T)}/(\pi k_B T_c)}
\label{eq:inducedgap}
\end{equation}
where  $\Delta_{\text{Al}}(T)$ is the BCS-like temperature dependence of the parent superconductor, and $\gamma_B$ parametrizes the transparency between epitaxial Al film and 2DEG ($\gamma_B=0$ for perfect vertical transparency, the larger $\gamma_B$, the larger the barrier). $\Delta^{\ast}(T)$ is calculated from Eq.~\ref{eq:inducedgap} and then plugged in the CPR equation (Eq.~3 of the main text) to obtain the Josepshon inductance $L$.

In the main text, we first fit the zero-bias $L$ as a function of $T$ using two parameters, namely $\Delta_{\text{Al}}(0)$ and $\gamma_B$. In this case the fit nicely matches all the experimental points and we obtain $\Delta_{\text{Al}}(0)=180$~\textmu  eV and $\gamma_B=1.0$. 

Alternatively, we determine $\Delta_{\text{Al}}(0)$ from the critical temperature, which is in turn deduced from the temperature dependence of the differential resistance, shown in Fig.~\ref{fig:roft}(a). Notice that foot of the transition to the normal resistance value can be nicely described as an activated process, as displayed in the Arrhenius plot in Fig.~\ref{fig:roft}(b). The activation energy is approximately 68~K$\times k_B$, which is of the same order of magnitude as the barrier for the thermally activated phase slips, namely twice the Josephson energy, i.e.~2$\hbar I_c/(2ek_B)\simeq$~210~K~\cite{Tinkhambook}.

Our criterion for $T_c$ is $R(T_c)=0.5R_N$, where $R_N$ is the normal resistance. From the graph in  Fig.~\ref{fig:roft}(a) we deduce $T_c=1.44$~K, which implies (assuming here a BCS relation) $\Delta_{\text{Al}}(0)=1.764 k_B T_c= 220$~\textmu  eV.  With $\Delta_{\text{Al}}(0)$ fixed, our fit procedure has only one parameter left, namely $\gamma_B$ (best fit $\gamma_B=1.7$). In this case, the model cannot suitably fit the data over the whole temperature range. A possible explanation for the discrepancy could be the non-perfect correspondence between $1.764 k_B T_c$ (with the above definition for $T_c$) and $\Delta_{\text{Al}}$. Another possibility is that the deviation at high temperature stems from the tails of the upward kinks of the $L(I)$ graphs in Fig.~2 of the main text.

Either way, we deduce a low temperature-limit for the induced gap $\Delta^{\ast}(0)\approx 130$~\textmu  eV. In fact, $\Delta^{\ast}(0)$ is mostly independent of $\gamma_B$ since its value is determined by the lower temperature regime, where the temperature dependence in Eq.~3  is mainly given by the $(1/2k_BT)$ factor in the $\tanh$ function, while $\Delta^{\ast}$ is still roughly temperature independent.

%{\red This part will be 100 percent settled when we will agree to a value for $T_c$. Then, $\Delta^{\ast}(0)$ will also affect N(0), see below. Also, I would add here a R(T) graph.}.

%In Fig.~2(e) of the  main text,  we fit the measured temperature dependence of the zero-bias Josephson inductance $L(0)$ with the corresponding values computed from the short-ballistic CPR (main text, Eq.~3), with $\Delta^{\ast}(T)$ deduced from Eq.~\ref{eq:inducedgap} above.

%In one case we keep free both $\gamma_B$ and $\Delta_{\text{Al}}$, obtaining perfect matching with the data for $\gamma_B=1$  and $\Delta_{\text{Al}}=182$~\textmu  eV. Alternatively, we hold $\Delta_{\text{Al}}=1.764k_BT_c$, with $T_c=1.35$~K determined from a $R(T)$ measurement of the array. The resulting one parameter ($\gamma_B=1.64$) fit does not produce a perfect agreement with the experimental data anymore. This possibly indicates that the relation between $T_c$ measured from $R(T)$ and $\Delta_{\text{Al}}$ is only approximately correct. In any case, the resulting $\Delta^{\ast}(0)$ is approximately 130~\textmu  eV.

\section{Averaging}

In the system under study, we need to distinguish two ensembles. The first one is the set of transverse channels in the 3.15~\textmu  m-wide 2DEG, which carry current in parallel within each junction. The second ensemble is that of the 2250 junctions in series. We will discuss them separately.

\textit{Ensemble of transverse channels.} A ballistic 2DEG of finite width contains $N=$ transverse channels, each carrying 2$e^2/h$ units of conductance. When connected to two superconducting banks to form a SNS junction, each channel $i$ will have a transmission coefficient $\tau_i$, not necessarily the same. Let us assume that $g(\tau)$ is the distribution function of the $\tau$ coefficients, such that $\int_0^1 g(\tau) d\tau=N$.

The supercurrent carried by each channel is then $\iota_i=\iota_0 f(\tau_i,\varphi)$ (where $\iota_0\equiv e\Delta^{\ast}/\hbar$, thus $I_0=N\iota_0$), and the total current is $I=\sum_i^N \iota_i$. The question is then whether a $\bar{\tau}$ exists  such that $I=I_0 f(\bar{\tau},\varphi)$.
A first trivial possibility is that $g$ is very narrow around a certain $\bar{\tau}$. Alternatively, we can investigate the regime of small phases, such that the term $\sqrt{1-\tau\sin^2(\varphi/2)}\approx 1$. Then, the CPR is proportional to $\tau$.
$$\iota_i=\iota_0 f(\tau_i,\varphi)\approx \frac{\iota_0 \sin \varphi}{2}\tau_i,$$
therefore the total current is 
$$I=\int g(\tau^{\prime})\frac{\iota_0 \sin \varphi}{2}\tau^{\prime} d\tau^{\prime}$$
$$=\frac{I_0 \sin \varphi}{2}\frac{1}{N}\int g(\tau^{\prime})d\tau^{\prime}$$
$$\equiv\frac{I_0 \sin \varphi}{2} \bar{\tau},$$
where in the last line $\bar{\tau}$ is naturally defined as  $\frac{1}{N}\int g(\tau^{\prime})d\tau^{\prime}$. In the measurements here reported the phase is always relatively small (see, e.g., Fig.~2(c) of the main text), owing to the effect of the weak junctions which limit the maximum bias current. Therefore writing Eq.~3 with an average $\bar{\tau}$ is approximately correct, as long as the phase is not too large \textit{and} the distribution $g(\tau)$ is not too broad.  

\textit{Ensemble of junctions in the array.} In the previous paragraph we showed that for a generic junction (let us label it with the index $j$) we can identify a $\bar{\tau}_j$ so that the supercurrent can be written as $I=I_0 f(\bar{\tau}_j,\varphi)$. This CPR implies an inductance $L_j(I)$ given by Eq.~2 of the main text. The junctions in the array, however, might show different   $\bar{\tau}_j$ values. The total inductance is the sum of the individual inductance in series, $L(I)=\sum_j L_j(I)$. However, the fact that we observe a near-unity $\bar{\tau}=0.94$ indicates that the distribution of the transmission coefficients is quite narrow. In fact, a significant amount of junctions with low $\bar{\tau}_j$ (producing a larger curvature in the normalized graph in Fig.~2(d)) would require a similar amount of junctions with above-unity transmission, which is impossible. It is thus reasonable to assume a distribution of $\bar{\tau}_j$ with a width of a few percent units around 0.94, which is anyway close to the experimental uncertainty (see magenta curves in Fig.~2(d) of the main text).

It is important to stress that there are indeed few junctions which we called \textit{weak} with a reduced critical current and/or reduced transmission $\bar{\tau}_j$. They produce the kinks in Fig.~2(a) of the main text. However, as long as the bias applied is well below their reduced critical current values, such weak junctions are too few to significantly affect the total inductance. In this sense, having a large arrays helps to average over the imperfection of single junctions, which is a clear advantage of our approach.

\begin{figure}[h!]
\includegraphics[width=1\columnwidth]{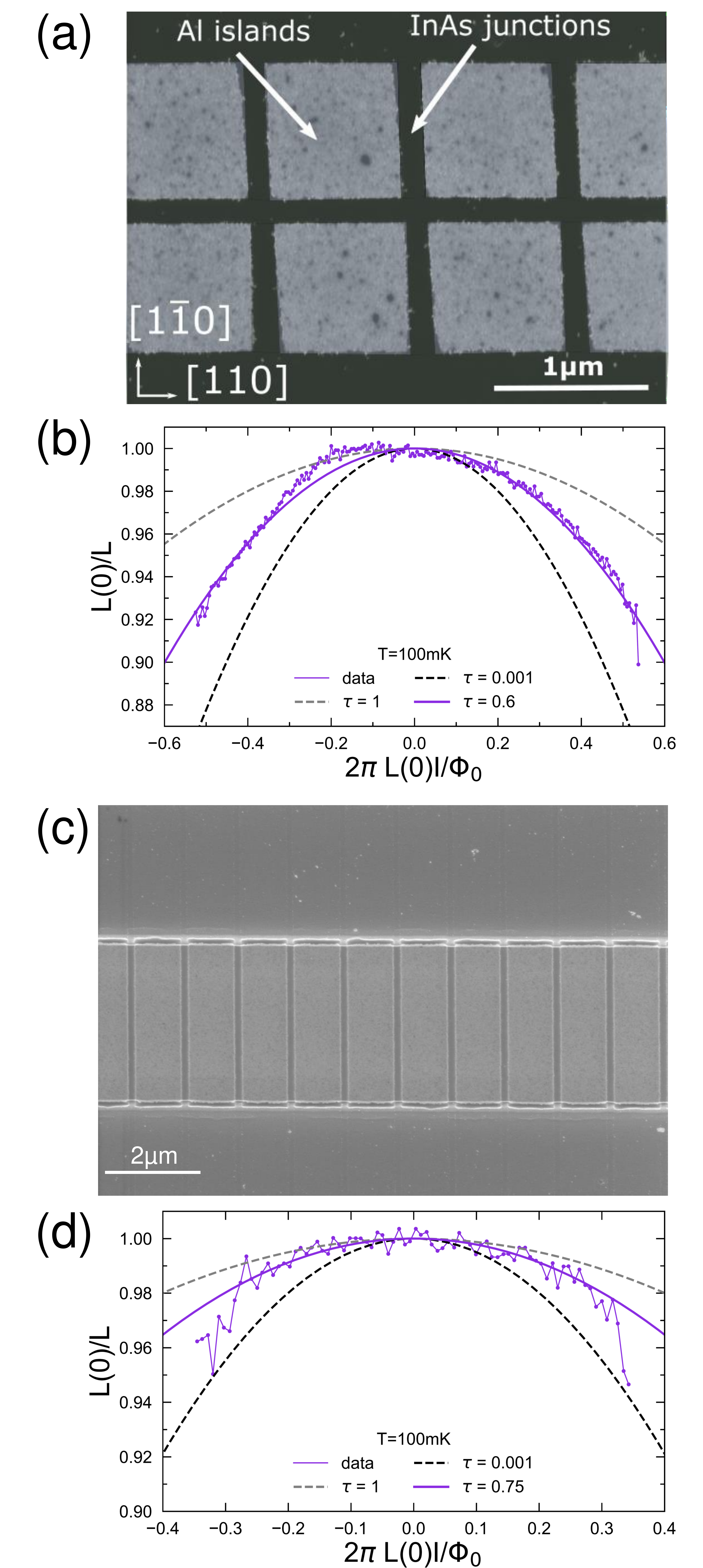}
\caption{(a) Scanning electron micrograph of Sample~1. (b) Plot of $L(0)/L$ versus $2\pi L(0)I/\Phi_0$ measured in Sample~1 at $T=100$~mK. The graph in this panel is analogous to that in Fig.~2(d) of the main text. (c) Scanning electron micrograph of Sample~3. (d) Plot of $L(0)/L$ versus $2\pi L(0)I/\Phi_0$ measured in Sample~3 at $T=100$~mK. 
}
\label{fig:others}
\end{figure}

\section{Other samples}

At the moment of writing we have completed the fabrication of several Josephson junction array devices similar to the one described in this article. Three of them (including the one described so far) were measured at low temperature and all of them showed a similar phenomenology.
The sample discussed in the main text is the one with the shortest gap between Al islands, and with highest transmission coefficient $\bar{\tau}$. In this section we discuss briefly the other two samples. In general, once the Al etching procedure has been optimized (by far the most delicate operation given how shallow the 2DEG is), then the fabrication procedure is in general reliable: if no evident lithographic defects are present, then the array works and displays a supercurrent over a large number of junctions.

Figures~\ref{fig:others}(a,c) show a scanning electron microscopy micrograph of Sample~1 and Sample~3, while the sample discussed in the main text is labelled as Sample~2. Notice that Sample~1 is an array with two rows of Al islands, but this does not have qualitative consequences at $B_{\perp}=0$. Figures~\ref{fig:others}(b) and (d) show the equivalent of the graph in Fig.~2(d) of the main text for Sample~1 and Sample~3, respectively. For the latter, also measurements at different temperatures are available. Sample~1 has a gap between Al islands of 150~nm and a width (sum of the two parallel islands) of 2~\textmu  m: its value of $L(0)$ at base temperature is $L(0)=345$~nH, with an average transmission coefficient $\bar{\tau}=0.6$.  Sample~3 has a gap between Al islands  which is not homogeneous (imperfect lithography), ranging from 130 to 180~nm, and a Al island width of 3.2~\textmu  m: its value of $L(0)$ at base temperature is $L(0)=217$~nH,  with an average transmission coefficient $\bar{\tau}=0.75$.

\end{document}